\documentclass[12pt,a4paper]{article}
\usepackage{babel,amssymb,amsmath,graphicx,hyperref}

\begin{document}

\begin{center}
{\Large Prospects for observing supermassive black hole binaries \\
with the space-ground interferometer}\\

\vskip8pt

{A.M.~Malinovsky$^{1*}$, E.V.~Mikheeva$^{1**}$}\\

\textit{$^1$ Astro Space Center of P.N. Lebedev Physical Institute of RAS,\\ Profsoyuznaya st., 84/32, 117997 Moscow, Russia}\\

\vskip5pt

$^*$amalin@asc.rssi.ru, $^{**}$helen@asc.rssi.ru

\end{center}

\vskip10pt

\begin{abstract}
A list of candidates for \textit{supermassive binary black holes} (SMBBHs), compiled from available data on the variability in the optical range and the shape of the emission spectrum, is analysed. An artificial neural network is constructed to estimate the radiation flux at 240~GHz. For those candidate SMBBH for which the network building procedure was feasible, the criterion of the possibility of observing the source at the \textit{Millimetron Space Observatory} (MSO) was tested. The result is presented as a table of 17 candidate SMBBHs. Confirmation (or refutation) of the duality of these objects by means of observational data which could be commited on a space-ground interferometer with parameters similar to those of the MSO will be an important milestone in the development of the theory of galaxy formation.
\end{abstract}

\section{Introduction}

It is believed that in the central part of any massive galaxy ($M>10^{12} M_\odot$) there is a \textit{supermassive black hole} (SMBH)\footnote{The SMBH is a black hole with mass $M>10^4 M_\odot$. The currently intensively investigated SMBHs have masses with $M>10^6 M_\odot$.}. This statement is now generally accepted, although unambiguous observational evidence for it is only available for the Milky Way  \cite{Gal1, Gal2}. For several hundred other galaxies, there are estimates of central SMBH masses measured by various methods \cite{catal1, catal2, catal3}. The most reliable methods include those based on the study of stellar or gas dynamics, while the currently widely used reverberation method contains a not well understood systematic error associated with the type of the object under study \cite{Huang2019}. Another new method of measuring the masses of SMBHs is the measurement of the black hole shadow using radio interferometry, which is not only an important achievement of observational astronomy, but also provides an independent estimate of the masses of SMBHs M87* \cite{M87} and SgrA* \cite{GC}.

Over the years of studying SMBHs in galaxy centers, several correlations have been found linking the mass of the central SMBH of galaxies with such parameters as the mass of the galaxy in which it is located, the mass of the stellar bulge, and the total mass of all globular clusters (see \cite{coev2, kormendy1997, tremaine2002} and references therein). The presence of such correlations undoubtedly points to a connection between the value of the SMBH mass and the evolution of galaxies (see \cite{coev1} and references therein). The nature of this relation can change with time, as indicated by recent observational data on the measurement of the masses of SMBHs and bulges of their host galaxies at high redshifts $z\simeq 6$ (see \cite{pens2020, tripodi2023}). 

The exact physical mechanism that gives rise the SMBH has not yet been elucidated. Of course, it is necessary to distinguish between the occurrence of "seeds" which could have a mass in the range of $10^2-10^5 M_\odot$, and the growth of the black hole mass due to accretion of matter from the surrounding space and/or merger with other black holes. According to the modern ideas, the SMBH seeds can arise as a result of 1) the direct collapse of a gas cloud, 2) the evolution of a dense stellar cluster, or 3) the merger of many stellar-mass black holes. \cite{volonteri2010}.

According to the modern concepts, the gravitationally bound dark matter halo, which is the dynamically dominant component of any galaxy, undergoes multiple mergers with halos of smaller masses during its evolution, and also in numerical simulations one can identify a main merger event when a halo interacts with another halo of close mass. This means that several SMBHs should be observed in galaxies for some time, and thus double (dual or binary) SMBHs \cite{volonteri2003}.

After capture of the SMBH by a halo containing another (more massive for certainty) black hole, the size of the SMBH orbit starts to decrease due to dynamical friction \cite{begelman1980}. The characteristic time of this process is $\sim 10^8$~years. At this time, SMBHs are gravitationally unbound and are referred to as "dual" SMBHs. When the distance between the SMBHs is reduced to $\sim1-100$~pc, a gravitationally-bound pair is formed, and the system of two SMBHs is starting to be "binary" \cite{rubinur2018}. It is such pairs of SMBHs are under consideration in the paper.

The further evolution of the SMBBH is determined by the dynamics of the pair's interaction with individual stars and the gas of the gravitationally-bound central star cluster, resulting in the loss of angular momentum and energy by the pair. Many uncertainties remain in the description of this stage, mainly related to the rate at the loss cone fills, making the duration of this stage hard to estimate. Nevertheless, when the distance between the SMBHs decreases to $10^{-2}-10^{-3}$~pc, the most efficient mechanism of energy loss by the the system becomes gravitational wave emission, leading to the merger of the SMBHs after $\sim10^8$~years. The details of the evolution of the binary system are an active area of research  \cite{review, roskar2014, volonteri2021}.    

A key characteristic of SMBBHs is that they are rare \cite{review}. Although the frequency of occurrence of binary systems remains uncertain and depends on the unknown evolutionary rate on small scales (the central parsec problem), the fraction of active galactic nuclei at redshift $z<0.7$ containing detectable SMBBHs is estimated to be $\sim 10^{-3}$ \cite{volonteri2009}. Close values were obtained using different approaches (see \cite{sesana2012a, kelley2019b, krolik2019}). 

Simple estimates show that the orbital period of the SMBH in the binary system, T, the large semi-axis, $a$, and the total mass of the SMBBH, $M+m$, are related by the relation
$$
\left(T/\text{year}\right)^2 \simeq 5.92\,\frac{\left(a/(0.01\, \text{pc})\right)^3}{(M+m)/10^9 M_\odot}.
$$
This means that, assuming to study the binary systems having periods not more than a few years, it is preferable to use sufficiently massive ("hypermassive", as proposed in \cite{catalmikh}, based on the capabilities of radio-interferometric observations) black holes separated by sub-pc distances. 

The best-known candidate for a SMBBH is the OJ~287 \cite{oj287}. According to the most preferred model, the mass of the main component is $\sim 10^{10} M_\odot$, the mass of the secondary component is $\sim 10^8 M_\odot$, and the orbital period in the binary system is $\sim12$ years. The large (as compared to other known SMBH) angular size of the more massive component shadow, $\sim 0.2\mu s$, and the total flux on the order of Jy make it a good candidate for observations with the space-ground interferometer \cite{ivanov2019}.

One of the main directions in search for SMBBHs is an analysis of the variability (in optics) of active galactic nuclei, mainly quasars. Thus, in \cite{catalina}, the \textit{Catalina Real-time Transient Survey} (CRTS) catalog containing 243500 spectroscopically confirmed quasars was analyzed. As a result, 111 objects showing signs of variability with periods on the order of a year were selected as candidate DSMBBHs. 

The search for the SMBBH candidates was also undertaken in \cite{palomar} by analyzing the \textit{Palomar Transient Factory} (PTF) catalog, which contains 35383 spectroscopically confirmed quasars. When analysing this catalog, 50 quasars with statistically significant periodicity were identified, and when analysing together with the CRTS catalog, 33 sources were added to the list of new candidates for SMBBH. Together with OJ~287, this make a total of 145 candidates for the SMBBH. 

The success of radio interferometry in the study of the SMBHs located at the center of galaxies M87 \cite{M87} and the Milky Way \cite{GC} suggests that similarly remarkable success can be achieved in the study of the SMBBHs. In other words, with a very high-resolution radio interferometer, it will be possible to see emission from the vicinity of both SMBHs of the binary system. However, flux data at 240 GHz are not readily available for the candidate SMBBHs selected for their variability in the optical band. This is partly due to the fact that observations in the sub-mm require s special conditions, firstly, a low concentration of water vapor in the atmosphere of the observation site. For this reason, it it logical to turn to modeling the flux values based on the available data.

The interest in SMBBHs is growing every year due to the important place of these objects in understanding the formation of SMBHs and host galaxies \cite{review}. So far, there is no unambiguous evidence for the duality of the available candidates even for such an intensively observed objects as OJ~287, i.e. very high-resolution interferometric observations are required.

In this paper, we present a methodology for modeling the spectra of SMBH in the mm and submm ranges using artificial intelligence methods. Using an artificial neural network, the values of the source fluxes at 240 GHz is used to test the observability criterion for sources on a space-ground interferometer with parameters similar to those of the MSO. The candidate SMBBHs satisfying the observability criterion are collected in Table~2. Since the available candidates were selected on the basis of properties indirectly related to the SMBH duality, i.e. may be due to other reasons, it is important to construct as complete catalog as possible for interferometric observations where the binary structure can be established.

\section{Modeling the spectra of SMBBH candidates} 

For the purposes of the study, out of the 145 known candidates for SMBBHs were selected those for which data on fluxes at higher and lower frequencies relative to the 240 GHz are available. There were 17 such objects. To this list from the literature were added 7 more candidate SMBBHs found in other publications (see Table~1). This table lists the name of the candidate sources, the mass estimate\footnote{The equality of component masses has been assumed.} expressed in units of the Sun mass, the physical distance between black holes in the binary system $D$ expressed in parsecs, the estimated value of the orbital period of the system $P$, the redshift $z$, and the reference.

\begin{table}[t] 
\caption{Candidates to the SMBBHs}
\begin{center} 
 \begin{tabular}{|l|c|c|c|c|c|} 
 \hline
 Source name & Mass,            & $D$, & $P$ & $z$ & reference\\
          & $\log(M/M_\odot)$ & pc   &     &     &      \\ 
 \hline
UM 269                   & 8.41  & 0.00313 & 490.5 days & 0.308 & \cite{palomar, catalina} \\
CSO 0402+379             & 8.18  & 7.3     & 150 000 years & 0.055 & \cite{mannes2004, rodriguez2006}   \\
FBQS J081740.1+232731    & 9.55  & 0.011   & 1190 days & 0.891 &  \cite{palomar, catalina} \\
BZQ J0842+4525           & 9.48  & 0.012   & 1886 days & 1.408 & \cite{palomar, catalina} \\
SDSS J084716.04+373218.1 & 8.1   & 0.022   & 40 years & 0.454 & \cite{liu2014, guo2019}  \\
OJ 287                   & 10.26 & 0.056   & 4380 days & 0.306 & \cite{oj287}  \\
MCG +11-11-032           & 8     & 0.0036  & 760 days & 0.036 & \cite{review}  \\
SBS 0924+606B            & 8.9   & 0.044   & 40 years & 0.295 &  \cite{liu2014, guo2019} \\
SDSS J094715.56+631716.4 & 9.22  & 0.014   & 1724 days & 0.487 & \cite{palomar, catalina}\\
SDSS J093819.25+361858.7 & 9.32  & 0.007   & 1265 days & 1.677 & \cite{palomar, catalina}\\
SDSS J100021.80+223318.7 & 9.3   & 0.052   & 35 years & 0.418 & \cite{decarli2010, eracleous2012p3}  \\
SDSS J102349.38+522151.2 & 9.59  & 0.014   & 1785 days & 0.955 & \cite{palomar, catalina}\\
SDSS J124044.49+231045.8 & 8.94  & 0.008   & 1428 days & 0.722 & \cite{palomar, catalina}\\
BZQ J1305-1033           & 8.50  & 0.008   & 1694 days & 0.286 & \cite{palomar, catalina}\\
SDSS J132103.41+123748.2 & 8.91  & 0.008   & 1538 days & 0.687 & \cite{palomar, catalina}\\
SDSS J133654.44+171040.3 & 9.24  & 0.008   & 1408 days & 1.231 & \cite{palomar, catalina}\\
SDSS J141244.09+421257.6 & 9.69  & 0.00622 & 433.4 days & 0.805 & \cite{palomar, catalina}\\
3C 298.0                 & 9.57  & 0.013   & 1960 days & 1.437 & \cite{palomar, catalina}\\
TEX 1428+370             & 8.53  & 0.00214 & 288.3 days & 0.566 & \cite{palomar, catalina}\\ 
SDSS J150243.09+111557.3 & 8.06  & 140     & 20 million years & 0.391 & \cite{deanenature2014}  \\
FBQS J150911.2+215508    & 8.54  & 0.00241 & 314.4 year & 0.438 & \cite{palomar, catalina}\\
PG 1553+113              & 8     & 0.0038  & 3 years & 0.360 & \cite{calieri2017}  \\
HS 1630+2355             & 9.86  & 0.020   & 2040 days & 0.821 & \cite{palomar, catalina}\\
PKS 2203-215             & 8.91  & 0.00408 & 497 days  & 0.577 & \cite{palomar, catalina}\\
\hline
\end{tabular}
\end{center} 
\end{table} 

An \textit{artificial neural network} (ANN) was built to model the flux at 240 GHz.

ANN is a mathematical model that has a structure similar to that of the brain, with "neurons" - computational units connected by synapses that communicate data between neurons. The ANN thus makes it possible to find the relationship between input and output data. 

Assume that the output data $y$ is a function of the input data $x$,
$y = f(x)$. In classical programming, the function $f$ is known, which makes it possible to determine the corresponding output data for given input data. For machine learning tasks, the function $f$ is usually not known. During the training of the model, it is provided with both input and output data, which makes it possible to establish the dependence between them. Then, while the trained model is running, the corresponding output data are calculated for the new input data.

In this case, the input data are frequency values and the output data are flux densities at these frequencies. First, the model was trained on publicly available data (see NASA/IPAC Extragalactic Database (NED)\footnote{https://ned.ipac.caltech.edu}). The trained network then determined the flux density at the frequencies of interest. \textit{Python} was used as a programming language and \textit{Tensorflow} as a machine learning library \cite{Tensorflow}.

In this work, a type of neural network called a \textit{multi layered perceptron} (MLP) is used, which is a forward propagation network, i.e. the signal propagates straight from the input of the network to the output. MLP consists of at least three layers: a layer (set) of input neurons receiving information, a hidden layer(s)\footnote{In our case, the network contained two hidden layers with 50 neurons each.}, which process the information, and a layer of output neurons, which output the results of calculations. When learning a network, the network uses what is called "supervised learning", i.e. the network is provided with many examples containing "known input - known output" pairs. At the beginning of the network training, the neuron weights (characterising their input) are activated randomly. By comparing the final result obtained with the given values and the known "output", it is possible to calculate the error, and then, using the backward propagation of errors (the gradient descent method), we can recalculate the weight values of all neurons. After that the new final result is calculated. The number of such cycles, called "epochs", was 10000, and the learning rate coefficient (a measure of the adjustment of weights in each epoch) was 0.01. As an activation function, which calculates output signal of the neuron depending on the sum of the inputs signals, was taken the so-called sigmoid, $f(x) = \frac{1}{1+e^{-x}}$.

The simulation results for all 24 sources from Table~1 are shown in Figures~1-3.

\begin{figure}
\begin{center}
\includegraphics[width=0.45\textwidth]{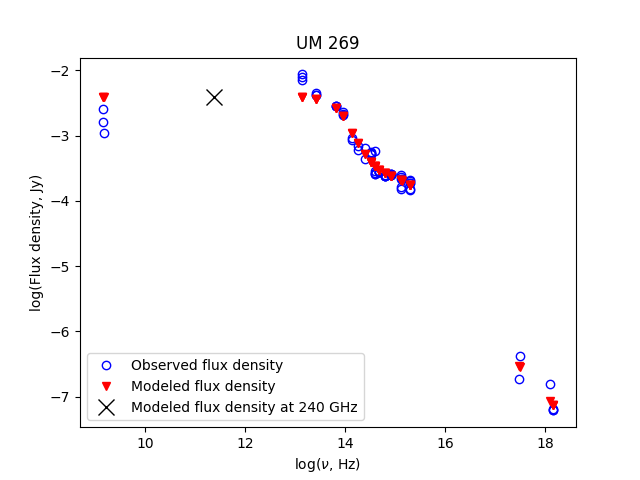}
\includegraphics[width=0.45\textwidth]{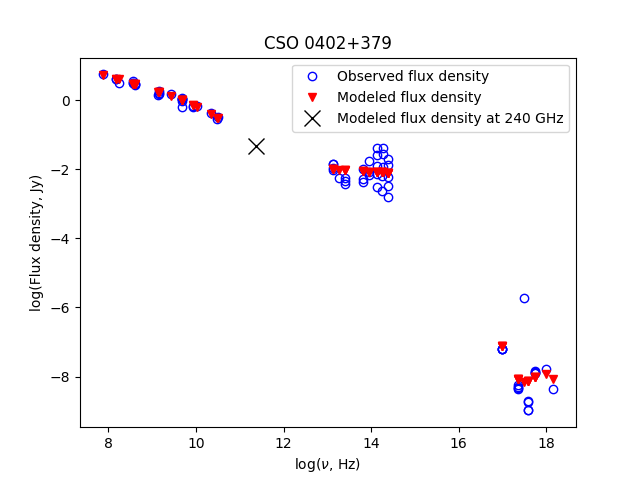}\\
\includegraphics[width=0.45\textwidth]{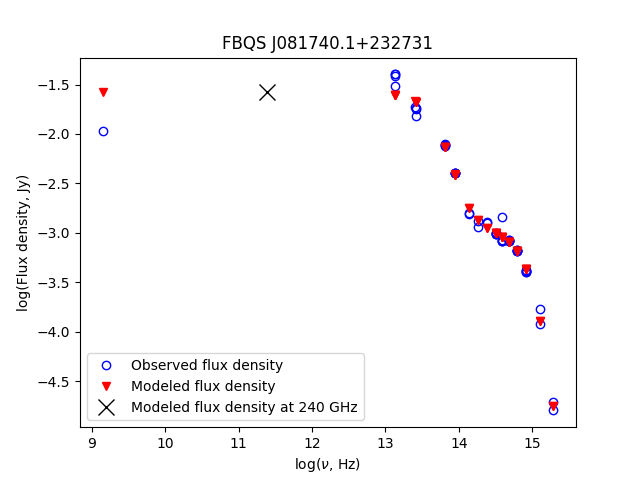}
\includegraphics[width=0.45\textwidth]{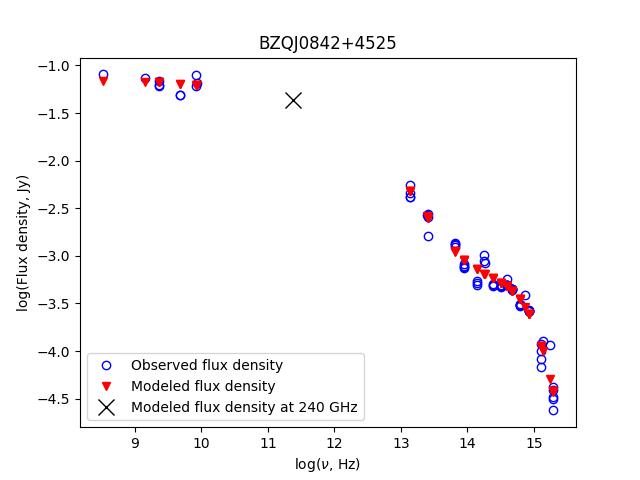}\\
\includegraphics[width=0.45\textwidth]{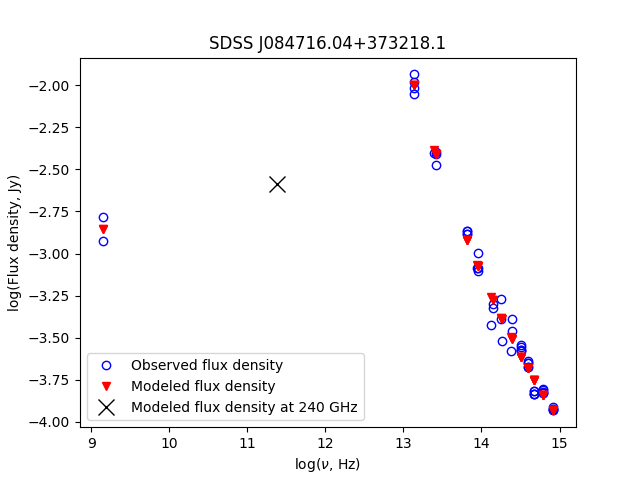}
\includegraphics[width=0.45\textwidth]{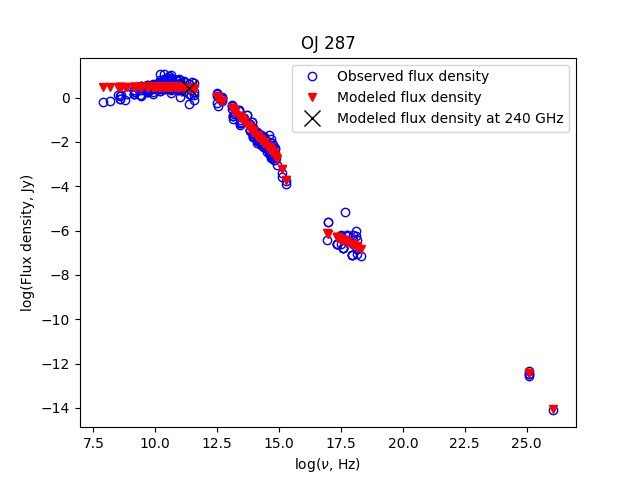}\\
\includegraphics[width=0.45\textwidth]{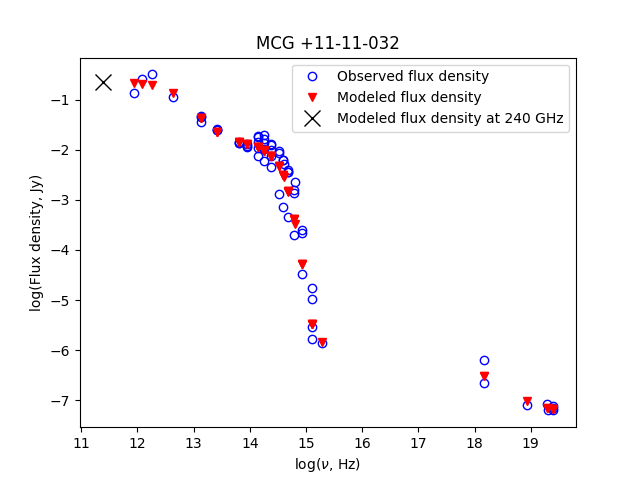}
\includegraphics[width=0.45\textwidth]{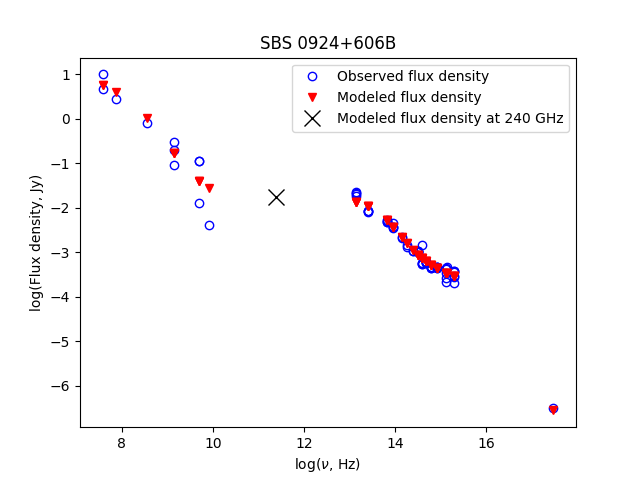}
\caption{Modeled spectra of the sources from Table~1. Part~1.}
\end{center}
\label{model1}
\end{figure}
\begin{figure}
\begin{center}
\includegraphics[width=0.45\textwidth]{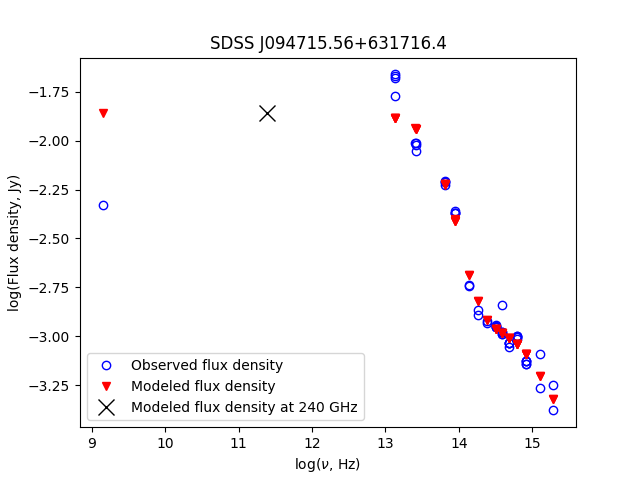}
\includegraphics[width=0.45\textwidth]{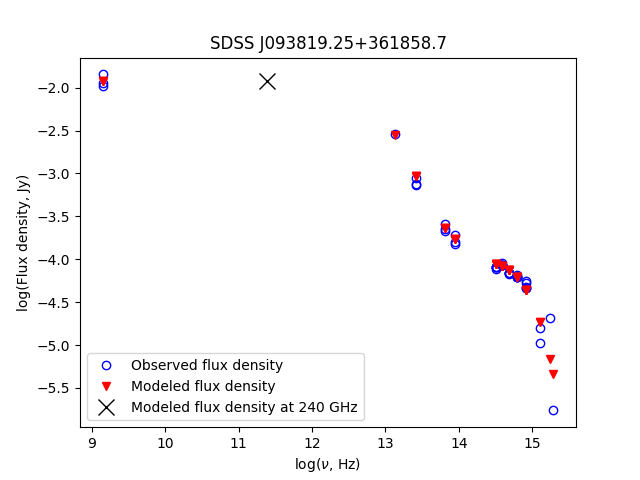}\\ 
\includegraphics[width=0.45\textwidth]{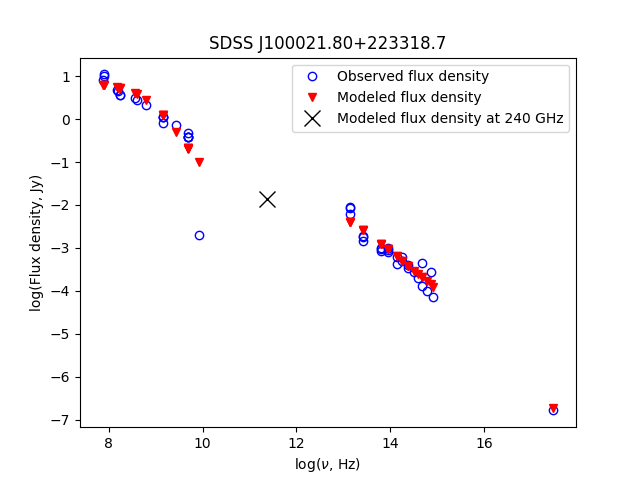}
\includegraphics[width=0.45\textwidth]{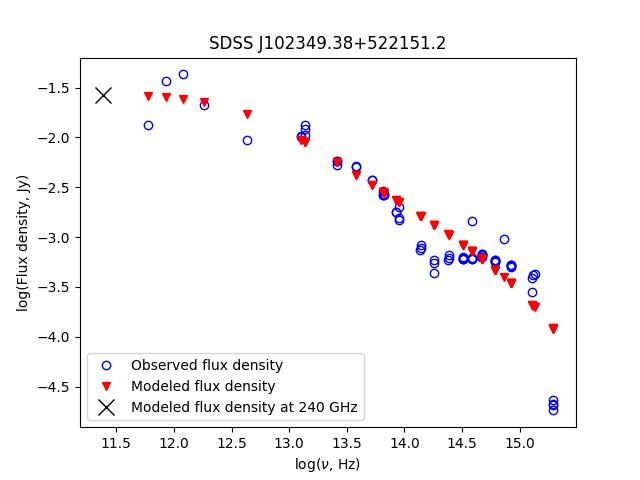}\\
\includegraphics[width=0.45\textwidth]{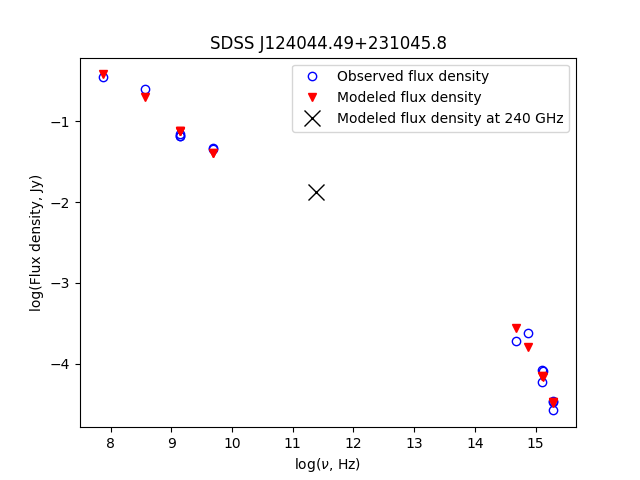}
\includegraphics[width=0.45\textwidth]{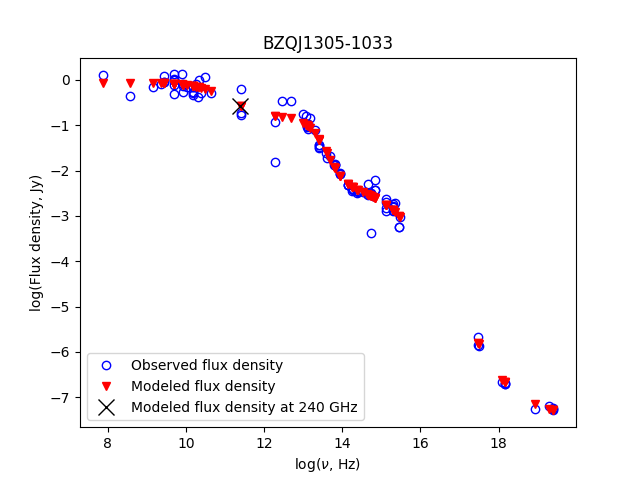}\\
\includegraphics[width=0.45\textwidth]{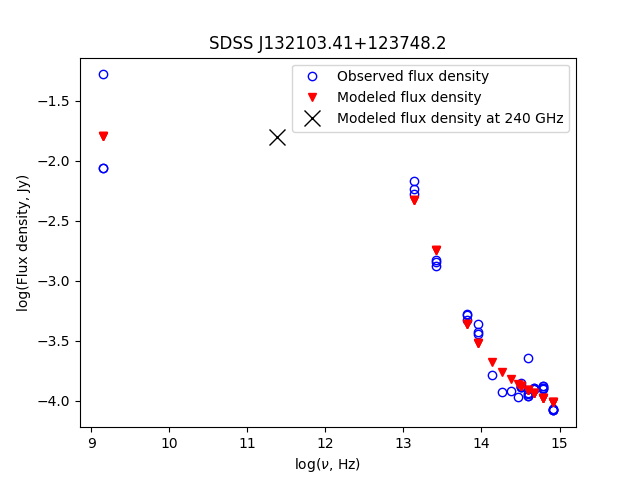}
\includegraphics[width=0.45\textwidth]{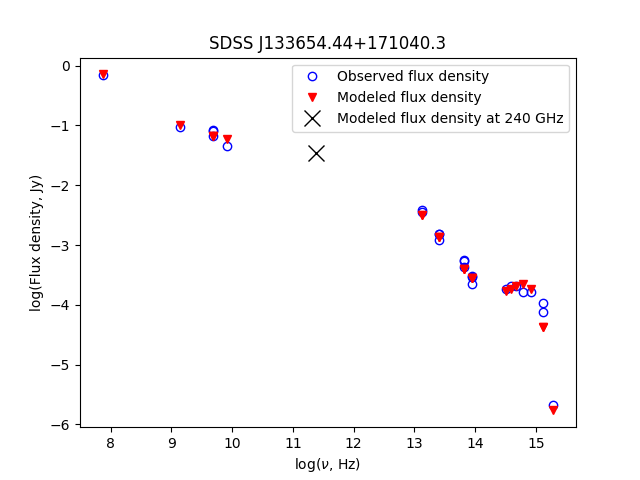}
\caption{Modeled spectra of the sources from Table~1. Part~2.}
\end{center}
\label{model2}
\end{figure}
\begin{figure}
\begin{center}
\includegraphics[width=0.45\textwidth]{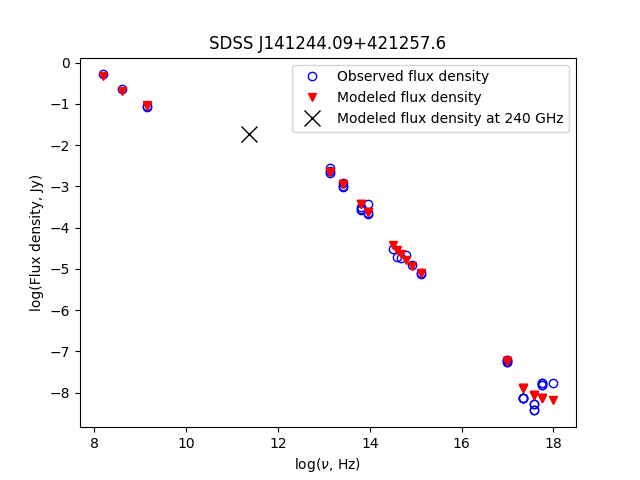}
\includegraphics[width=0.45\textwidth]{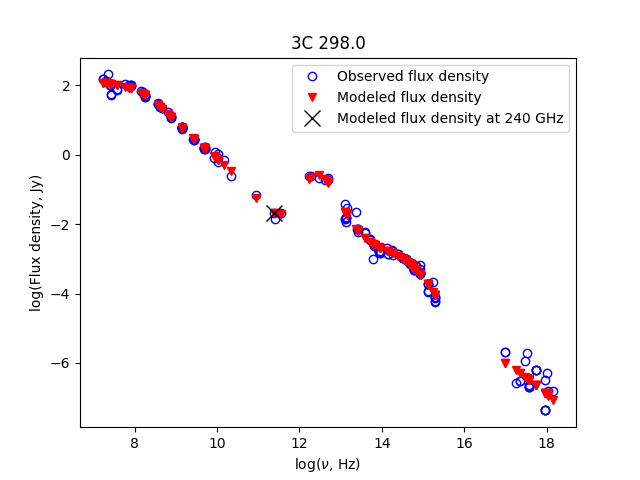}\\
\includegraphics[width=0.45\textwidth]{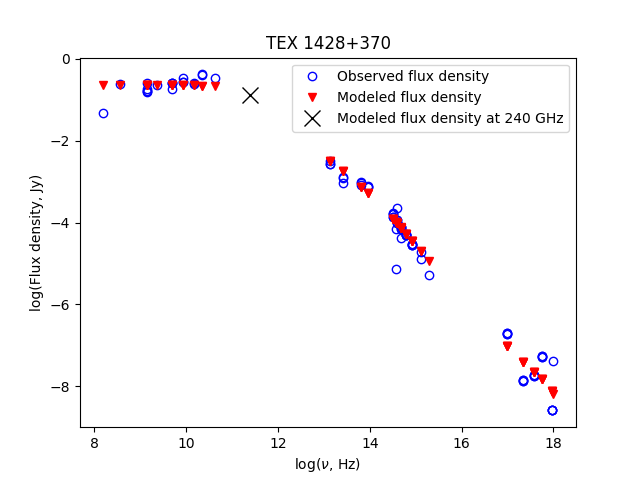}
\includegraphics[width=0.45\textwidth]{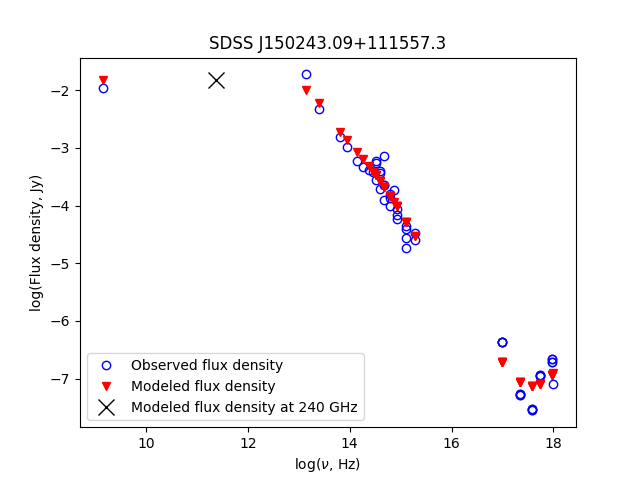}\\
\includegraphics[width=0.45\textwidth]{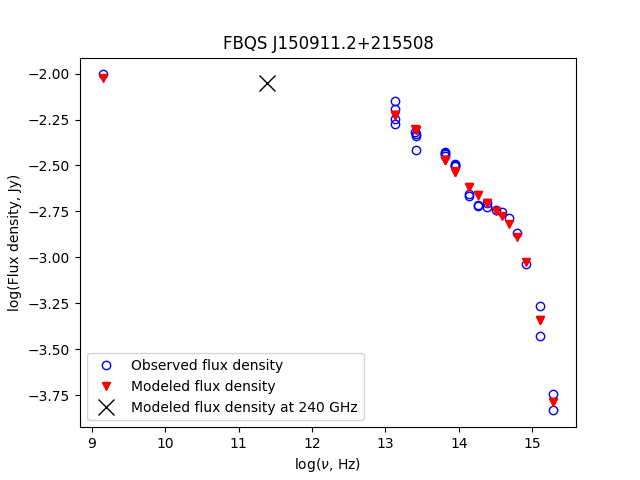}
\includegraphics[width=0.45\textwidth]{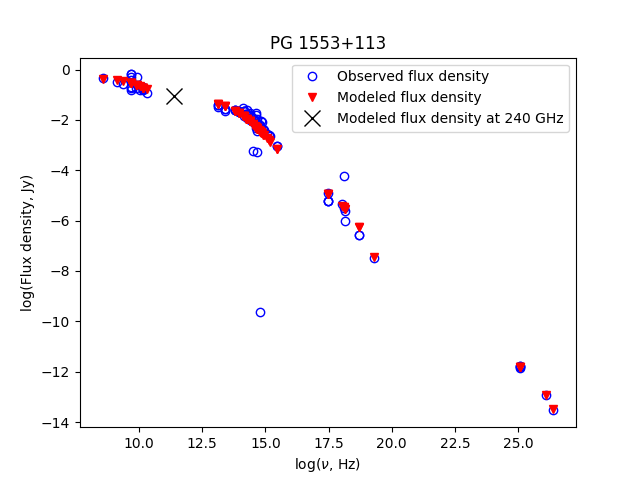}\\
\includegraphics[width=0.45\textwidth]{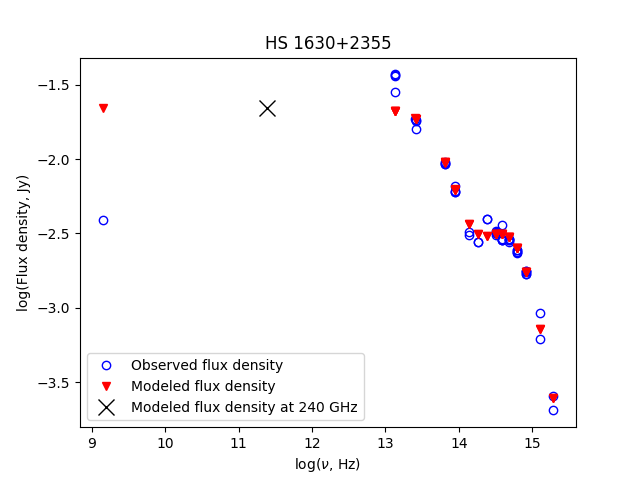}
\includegraphics[width=0.45\textwidth]{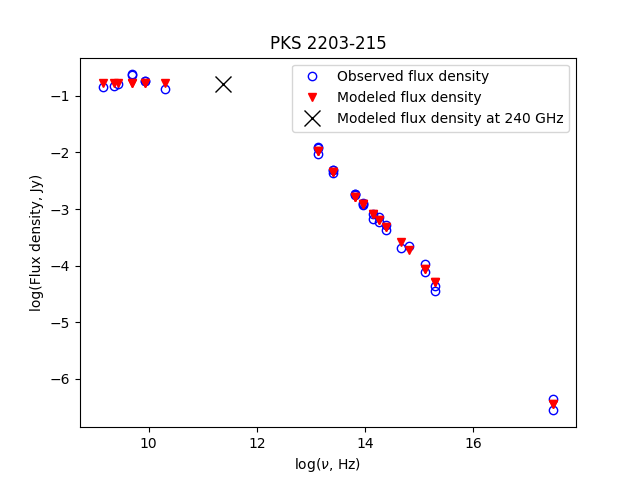} 
\caption{Modeled spectra of the sources from Table~1. Part~3}.
\end{center}
\label{model3}
\end{figure}

\section{SMBBH space-VLBI}

Since the angular sizes of SMBHs are very small, their space-ground VLBI observations in mm and submm range is of particular importance, see, for example, \cite{ivanov2019}. In the same paper, a source selection criterion for observations on the MSO was formulated. This criterion was adapted for the purposes of this study, with 240 GHz being used as the "main frequency" (see \cite{ivanov2019}).

According to the selection criterion, the modeled flux values obtained from the using ANN are substituted into eq.(8) from \cite{ivanov2019}:
$$
F_{\text{av}}=\sqrt{\frac{1}{2\pi} \int C(u,v)C^{*}(u,v)\text{d}\psi } =
\frac{F_{ANN}}{\pi\rho R(1-r^2)}\sqrt{J_1^2(x_1)+J_1^2(x_2)r^2-2J_1(x_1)J_1(x_2)J_0(x_3)r},
$$
where $F_{\text{av}}$ is the visibility function averaged over the azimuntal angle in the $u-v$-plane,  $\psi\equiv\arcsin(v/\rho)$, the value of $\rho$ is calculated from the possible values of the minimum and maximum projection of the bases for the selected orbit of MSO for the coordinates of the selected sources, $C(u,v)$ is the two-dimensional Fourier image of the source model, $J_\nu(x)$ is the Bessel function, $F_{ANN}$ is the modeled flux value, $R$ and $r$ are the parameters of the source model (a more detailed consideration of the source model can be found in \cite{ivanov2019}). Next, we checked whether there is a range of base projections in which the value of the averaged flux is higher than the detection limit, which is 6.45 mJy for 240 GHz for MSO. The dependence of the averaged flux for the selected sources on the value of the base projection, expressed in Earth diameters, is shown in Figs.~4-6. In all figures, the horizontal dashed line corresponds to the sensitivity level of the telescope. 

\begin{figure}
\begin{center}
\includegraphics[width=0.45\textwidth]{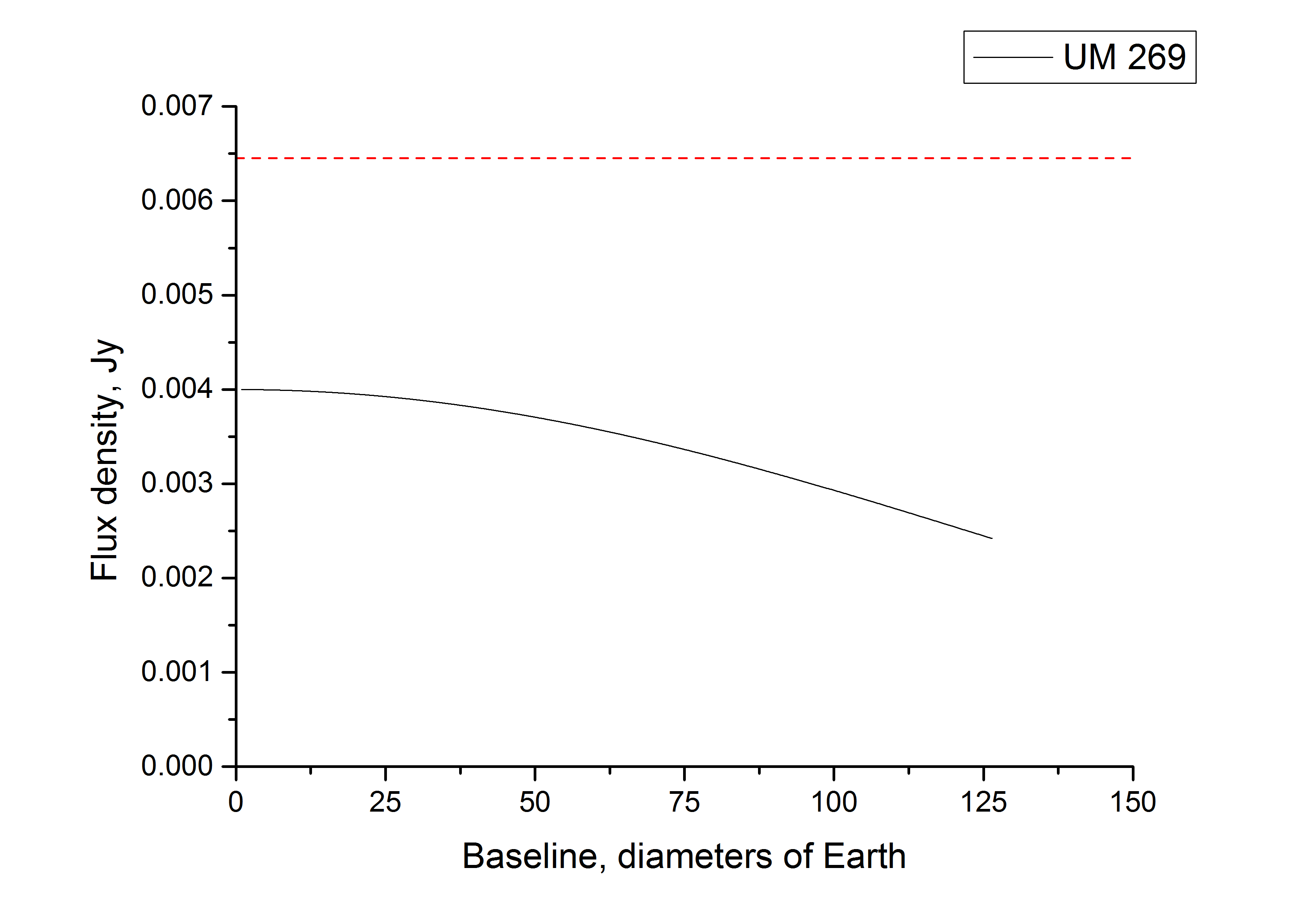}
\includegraphics[width=0.45\textwidth]{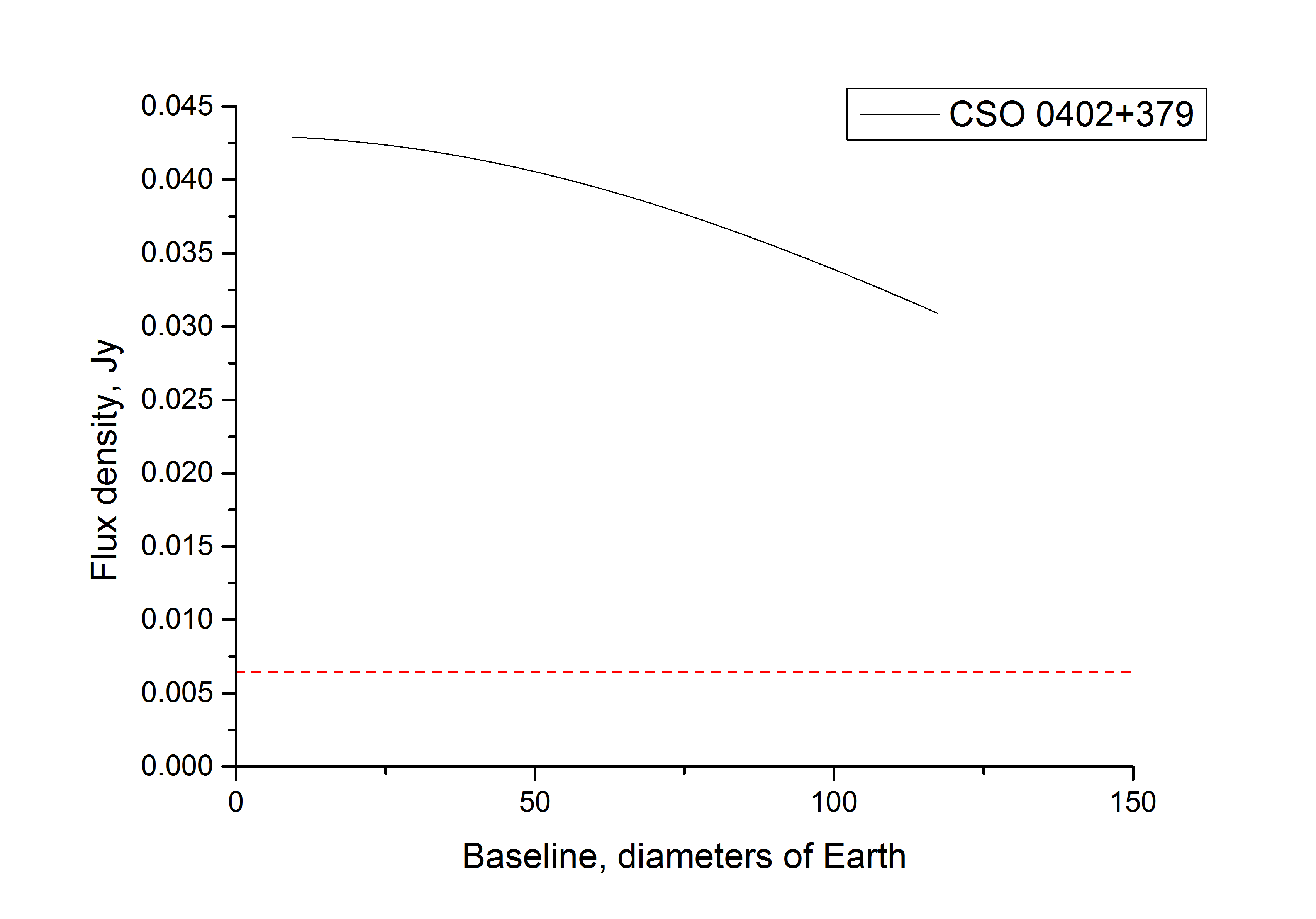}\\
\includegraphics[width=0.45\textwidth]{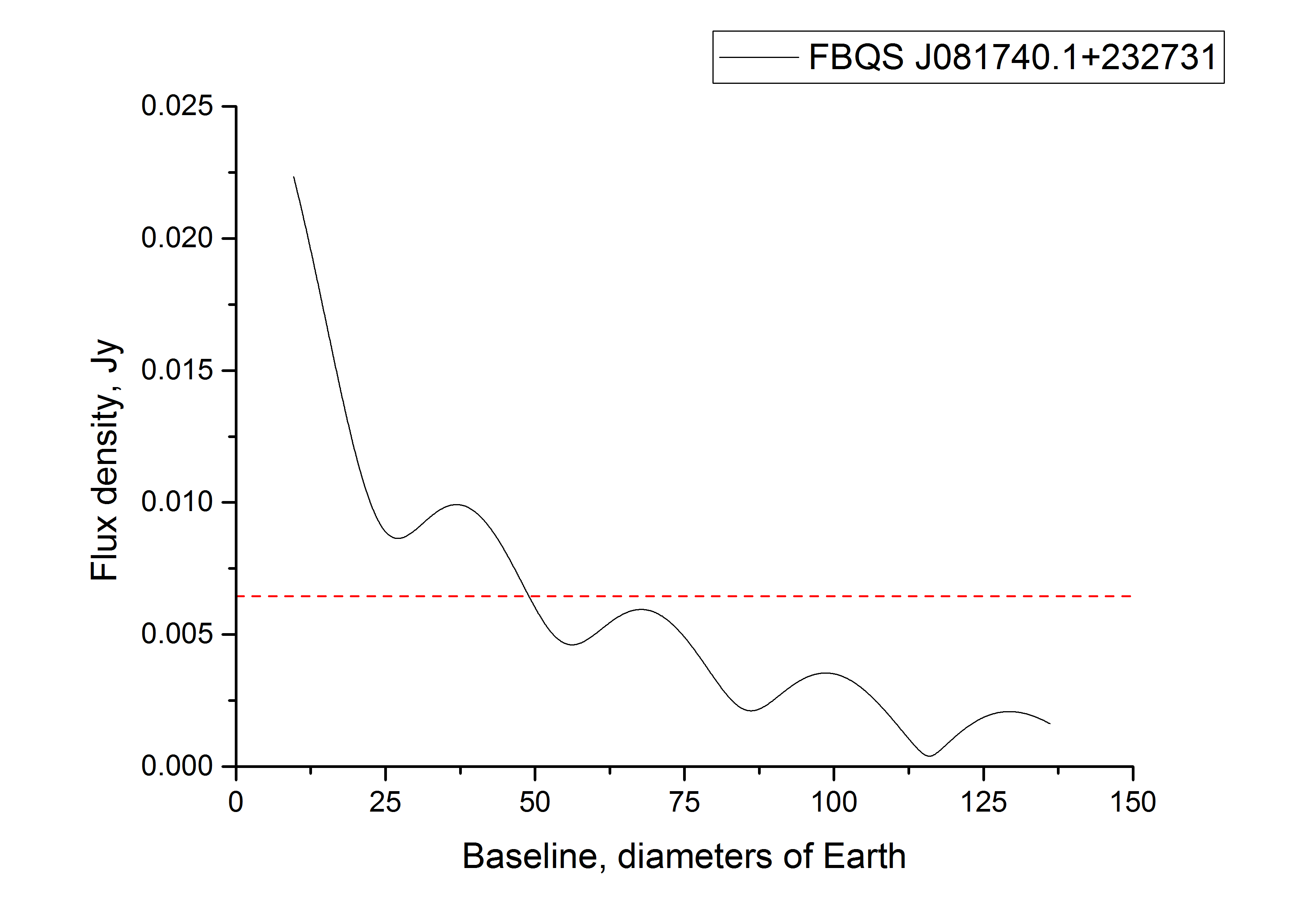}
\includegraphics[width=0.45\textwidth]{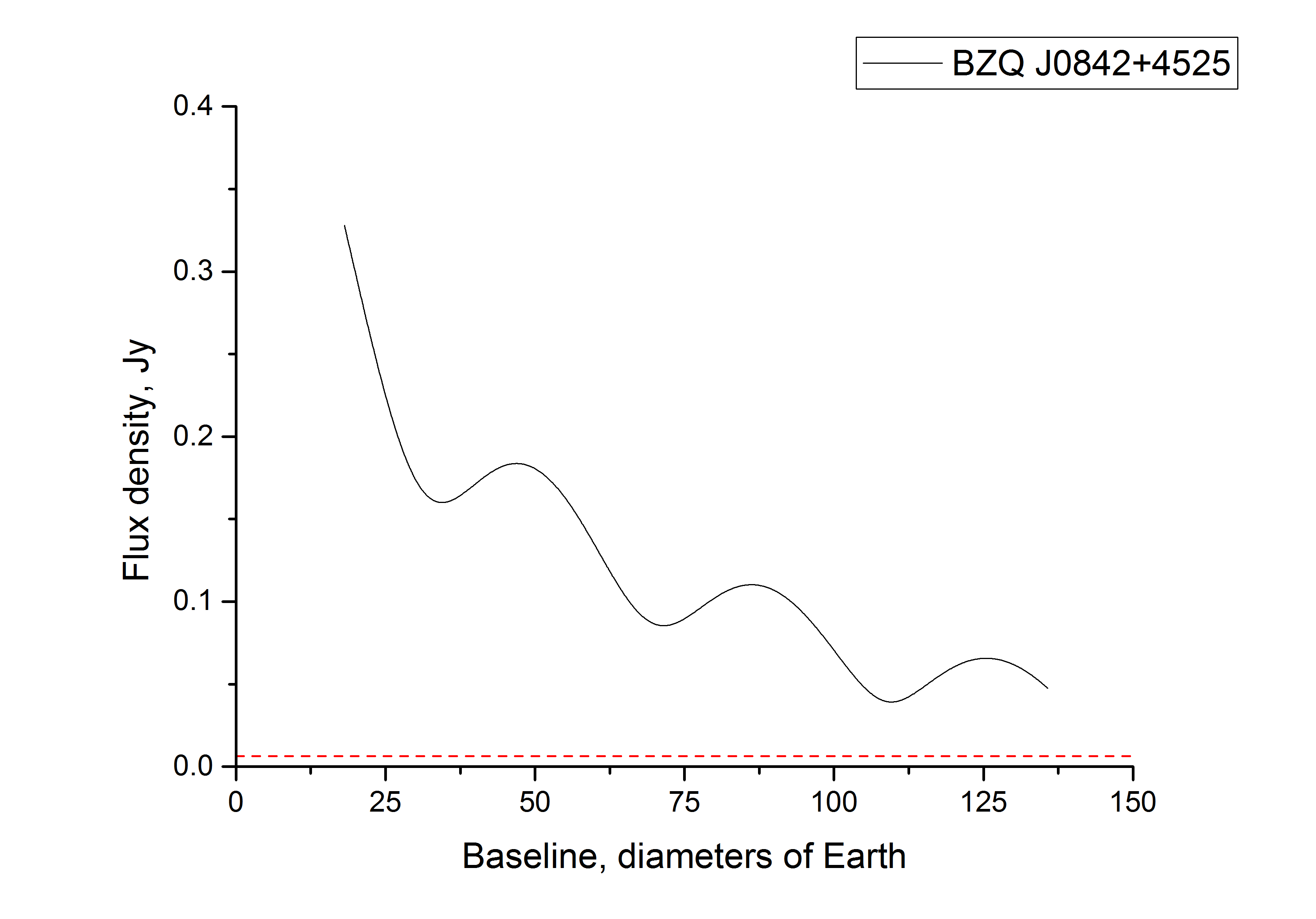}\\
\includegraphics[width=0.45\textwidth]{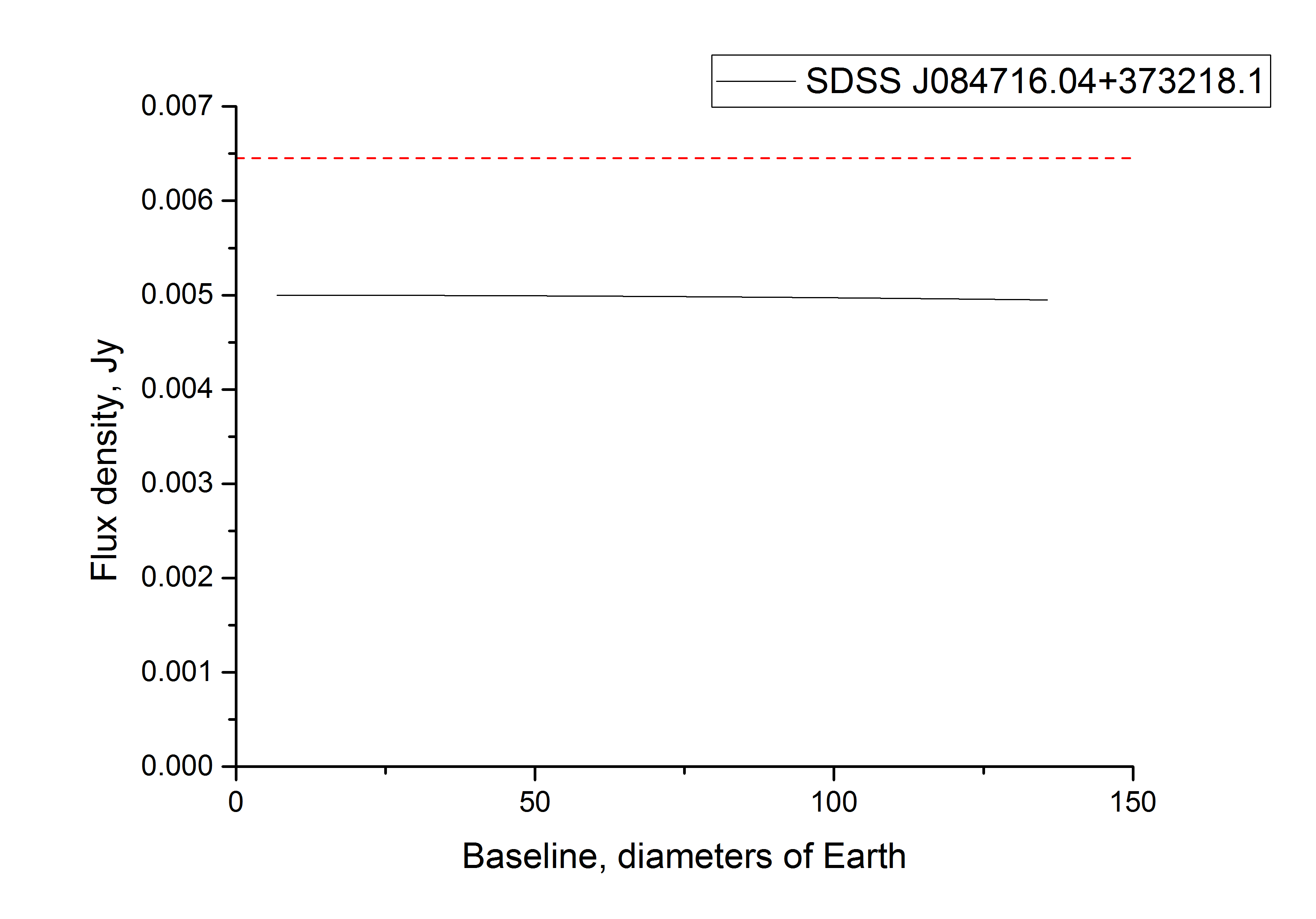}
\includegraphics[width=0.45\textwidth]{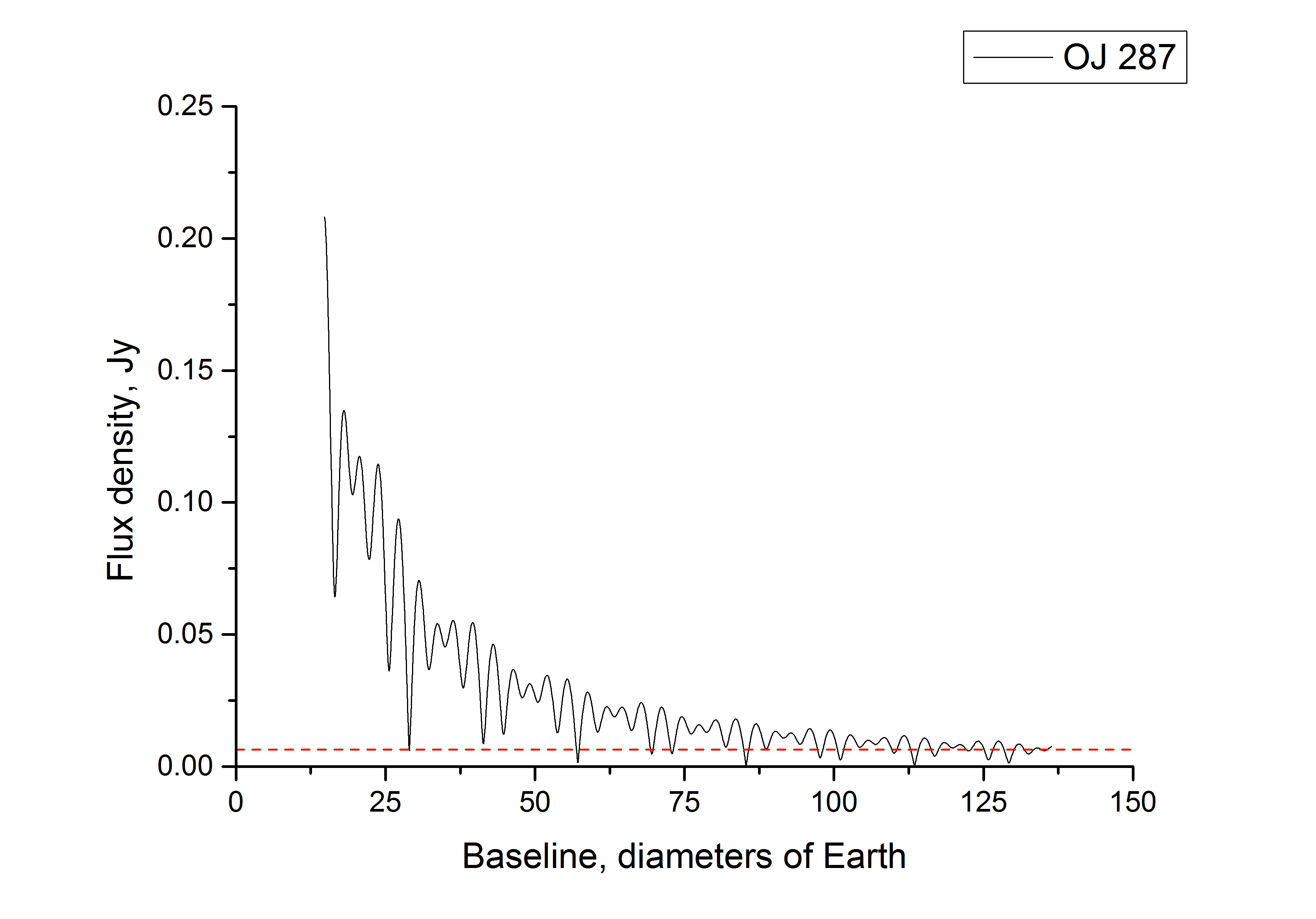}\\
\includegraphics[width=0.45\textwidth]{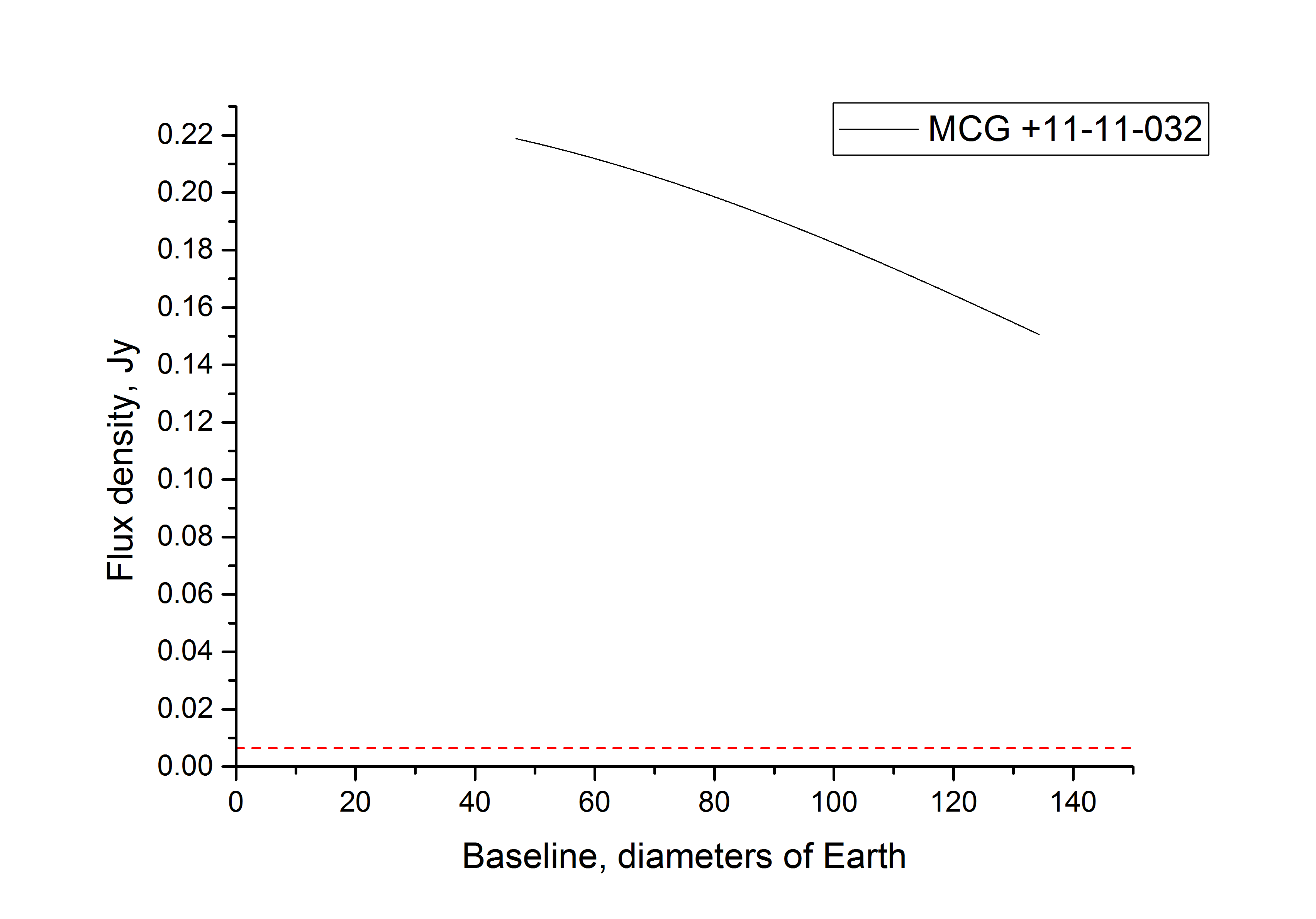}
\includegraphics[width=0.45\textwidth]{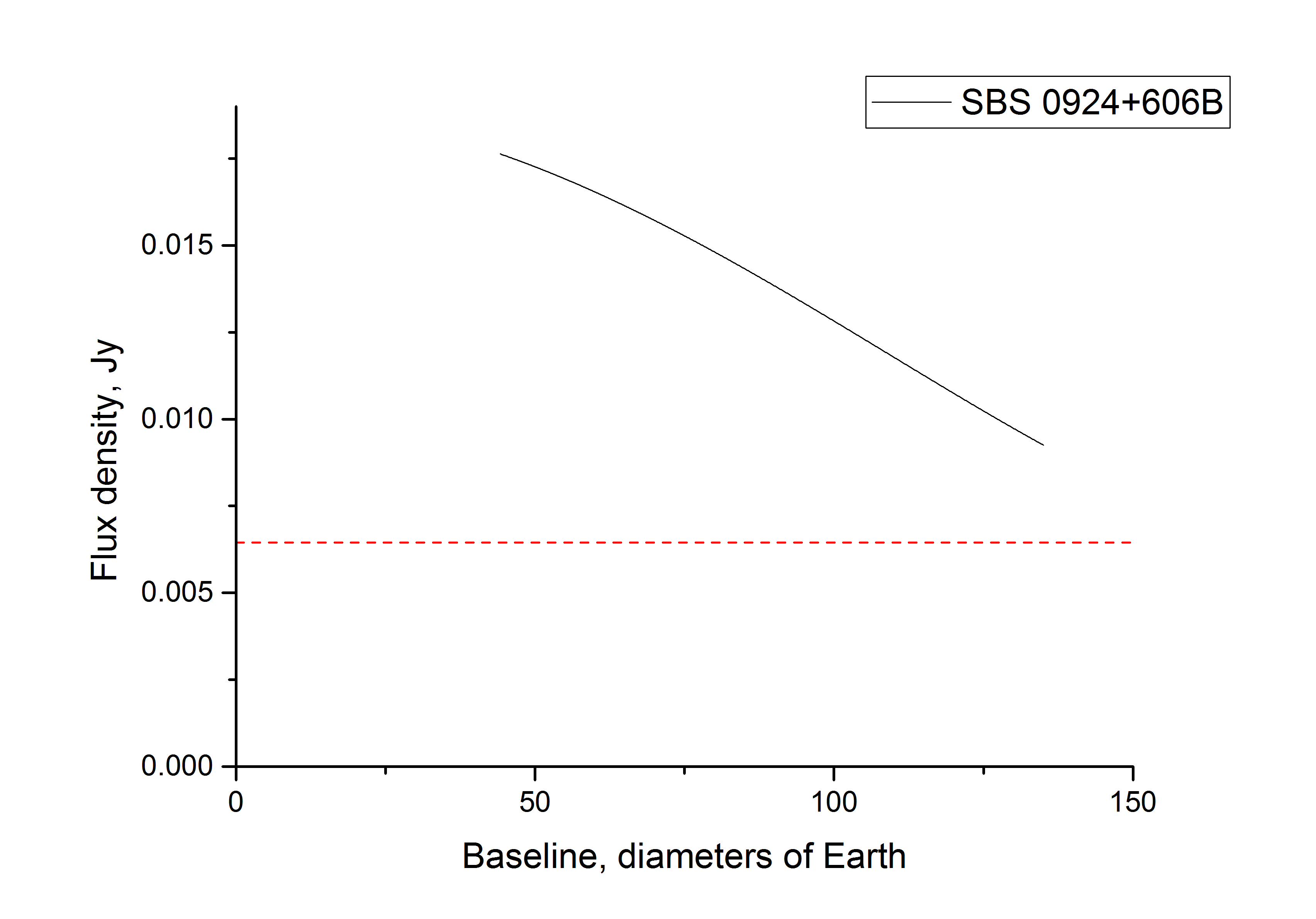}
\caption{Averaged visibility function for acceptable values of base projection for the sources from Table~1. Part~1.}
\end{center}
\label{flux1}
\end{figure}
\begin{figure}
\begin{center}
\includegraphics[width=0.45\textwidth]{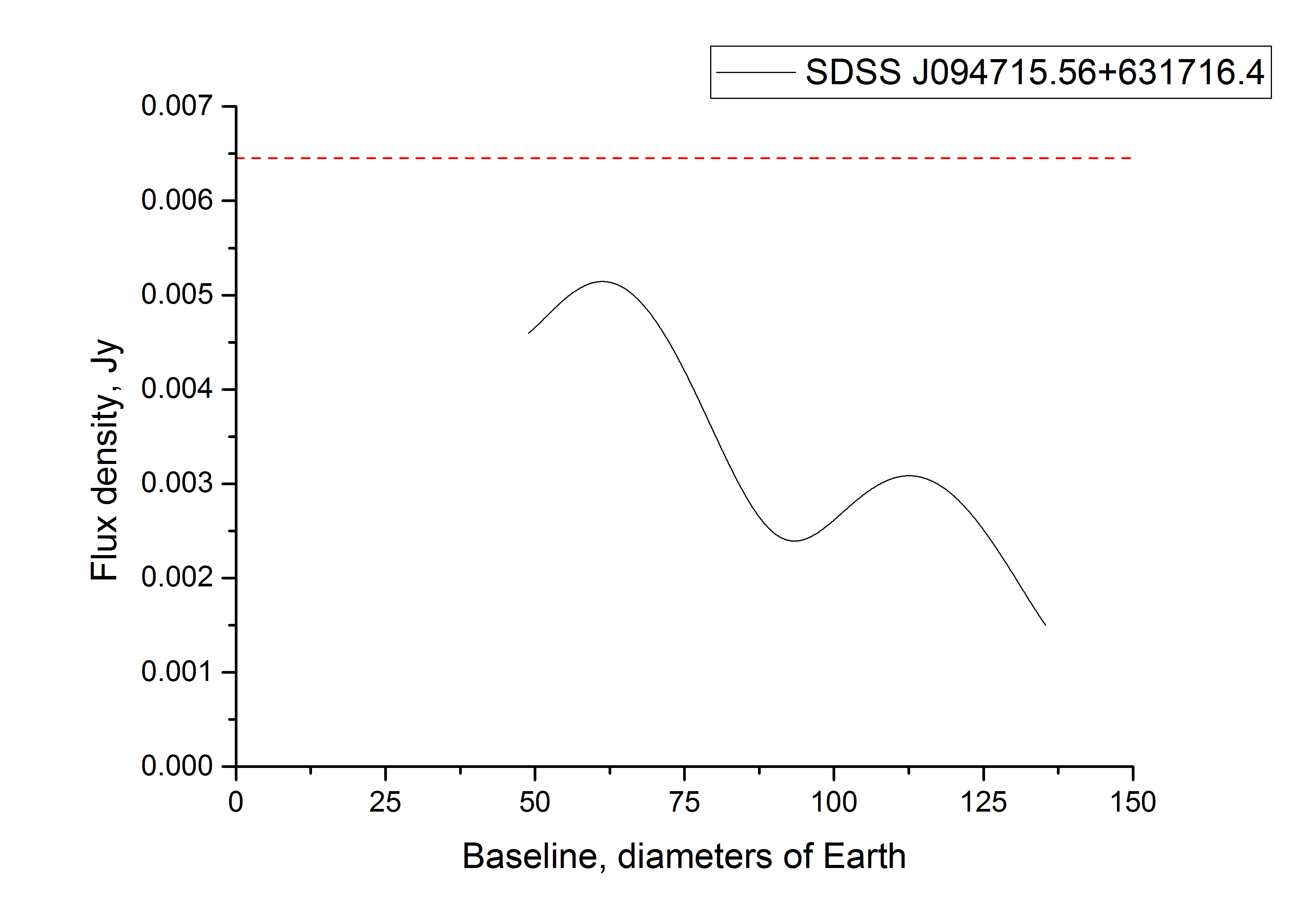}
\includegraphics[width=0.45\textwidth]{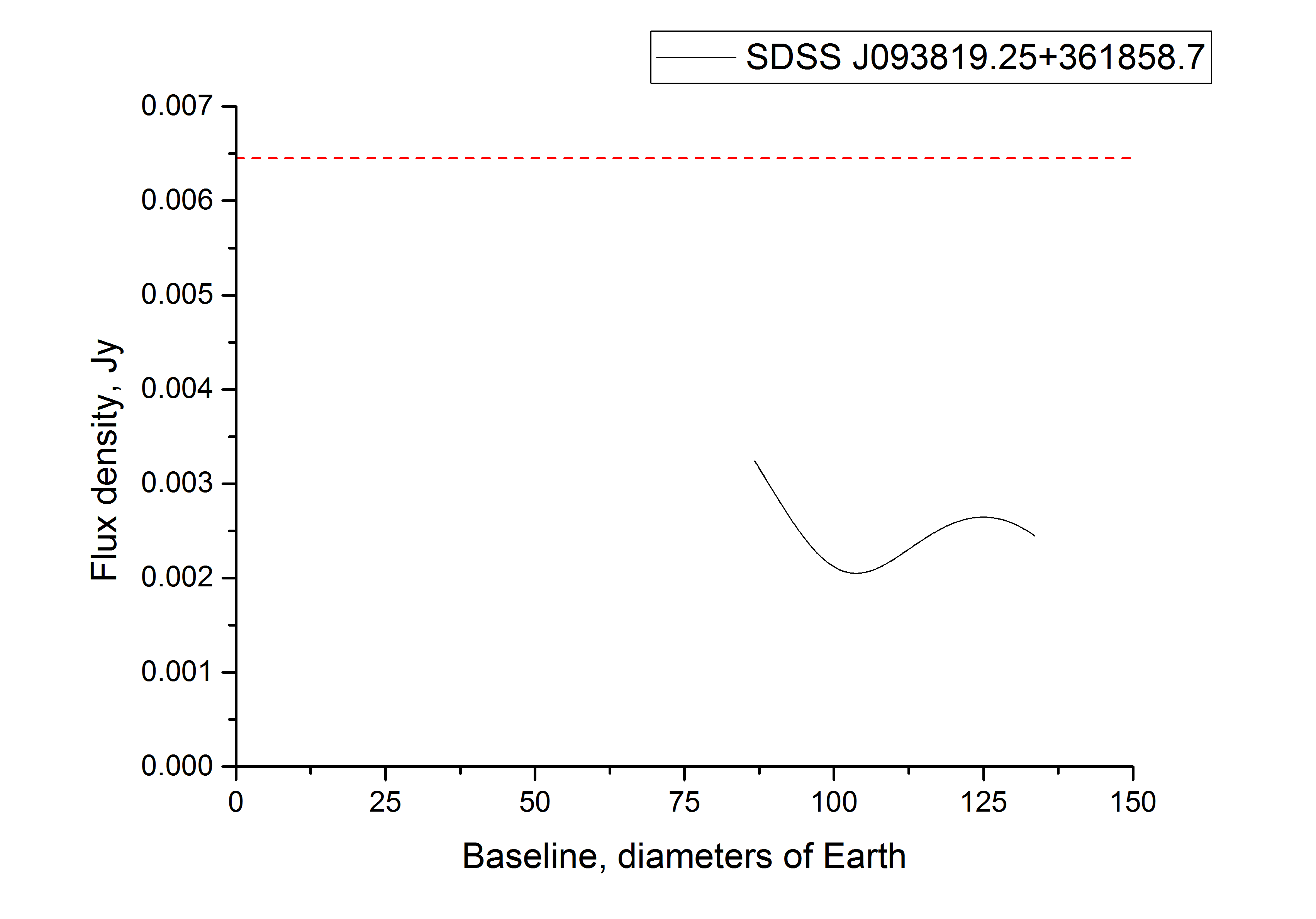}\\
\includegraphics[width=0.45\textwidth]{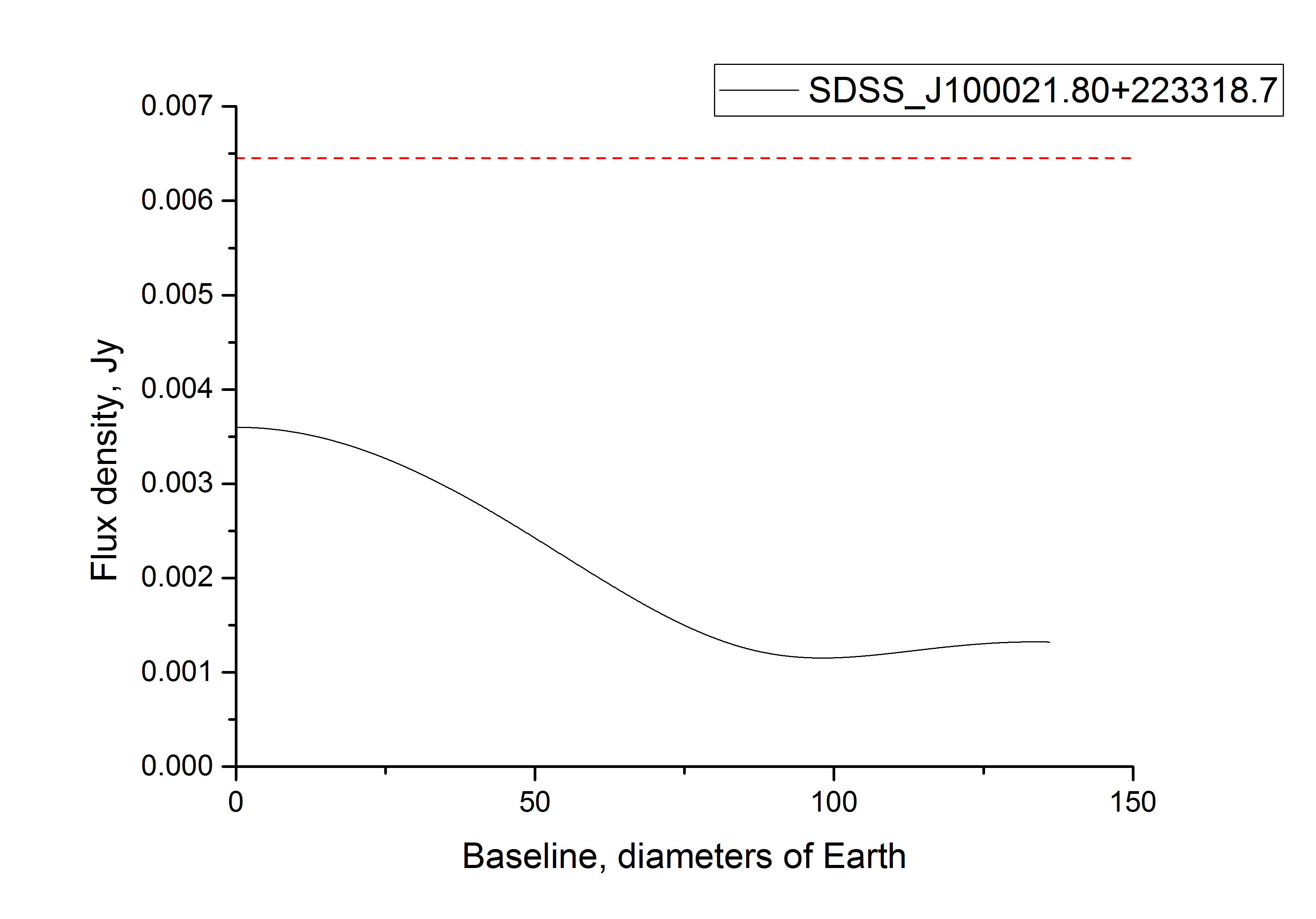}
\includegraphics[width=0.45\textwidth]{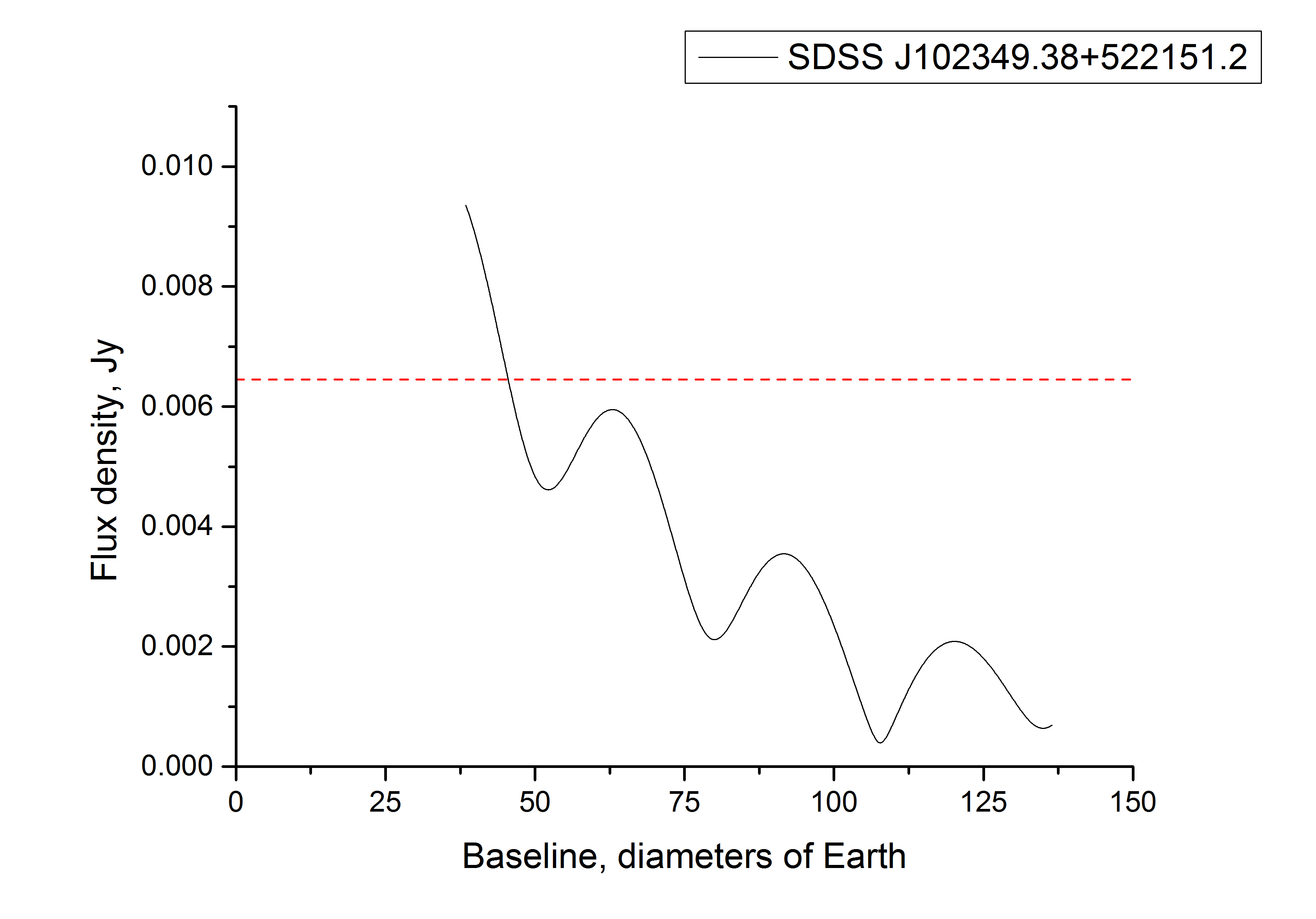}\\
\includegraphics[width=0.45\textwidth]{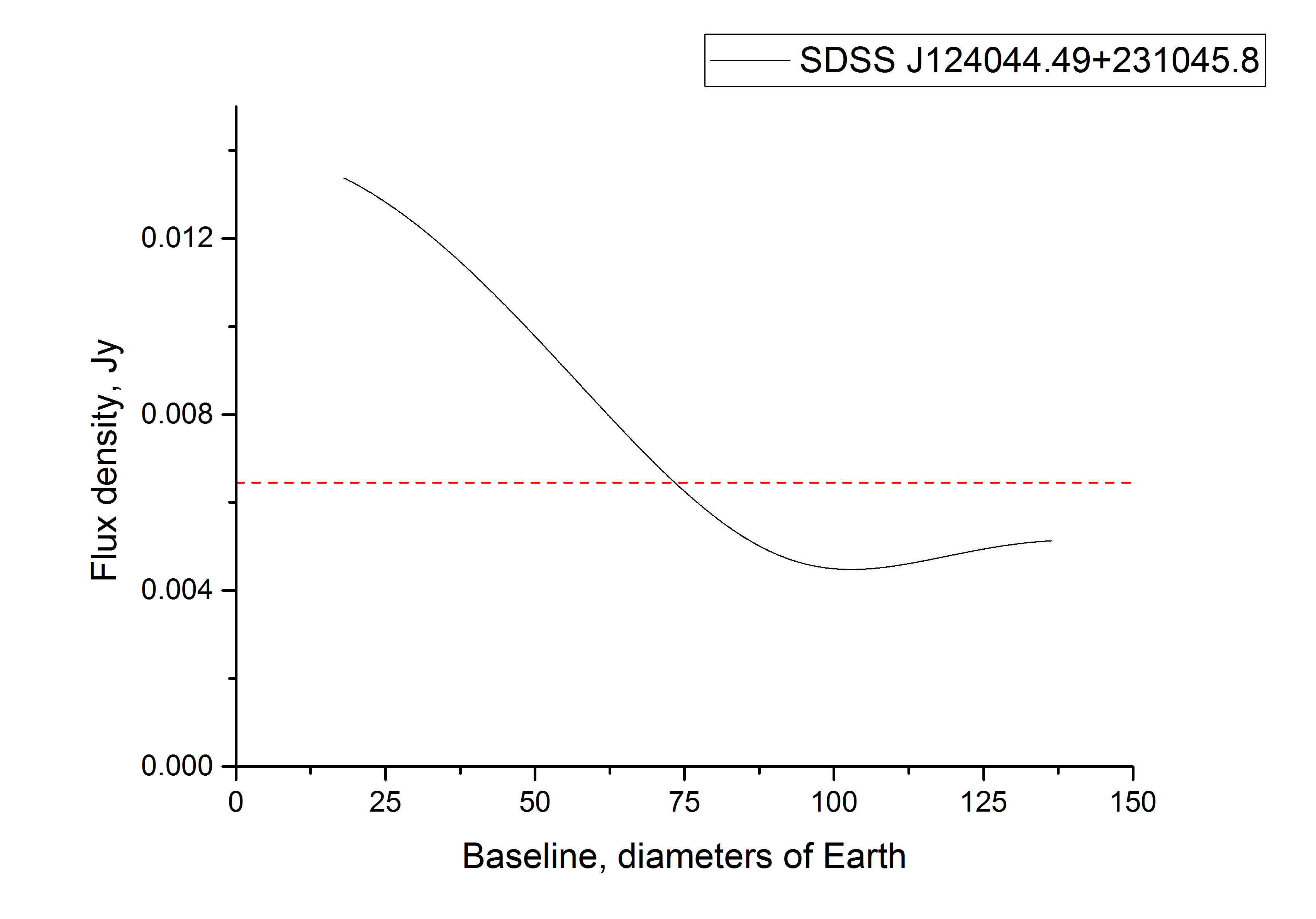}
\includegraphics[width=0.45\textwidth]{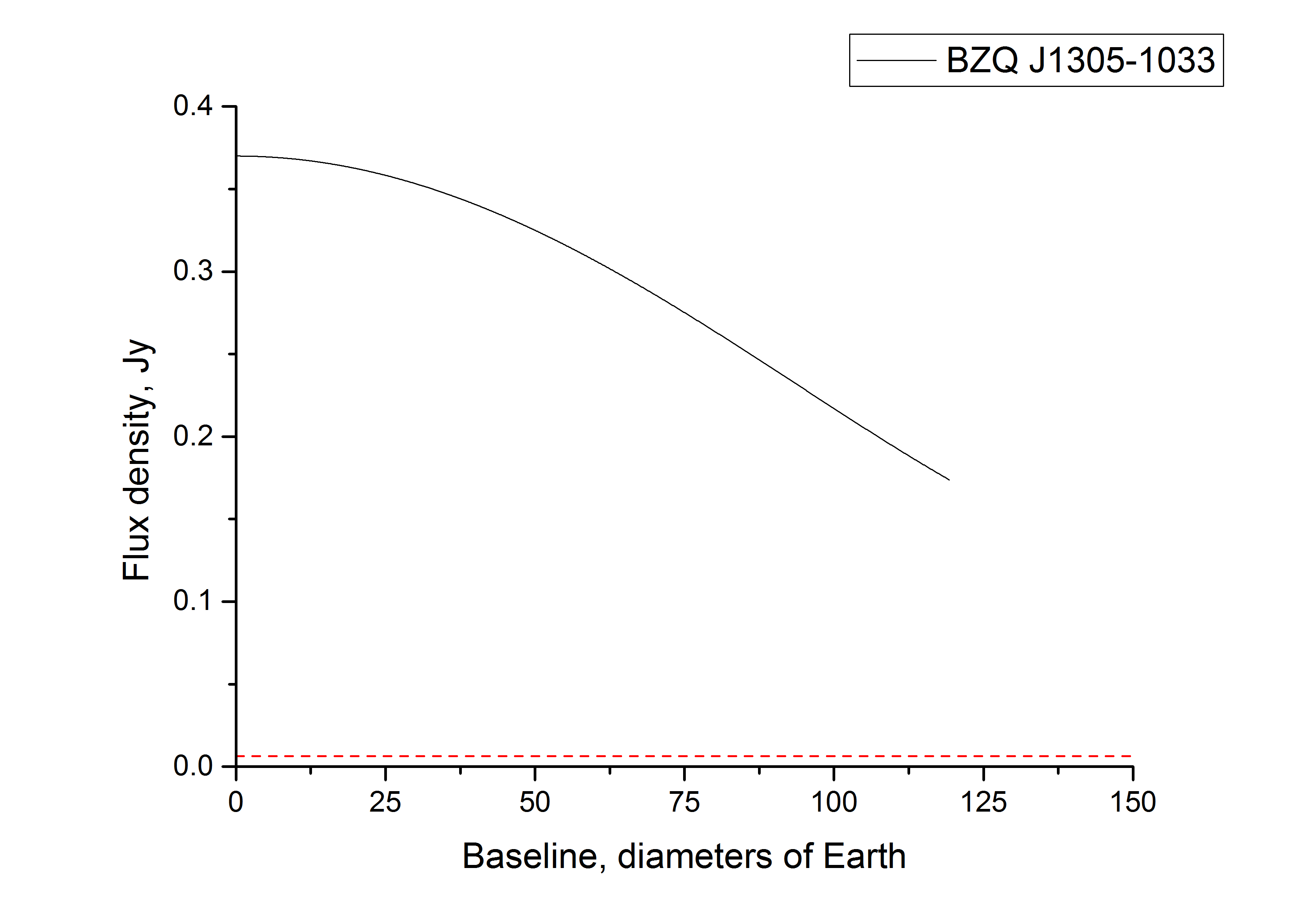}\\
\includegraphics[width=0.45\textwidth]{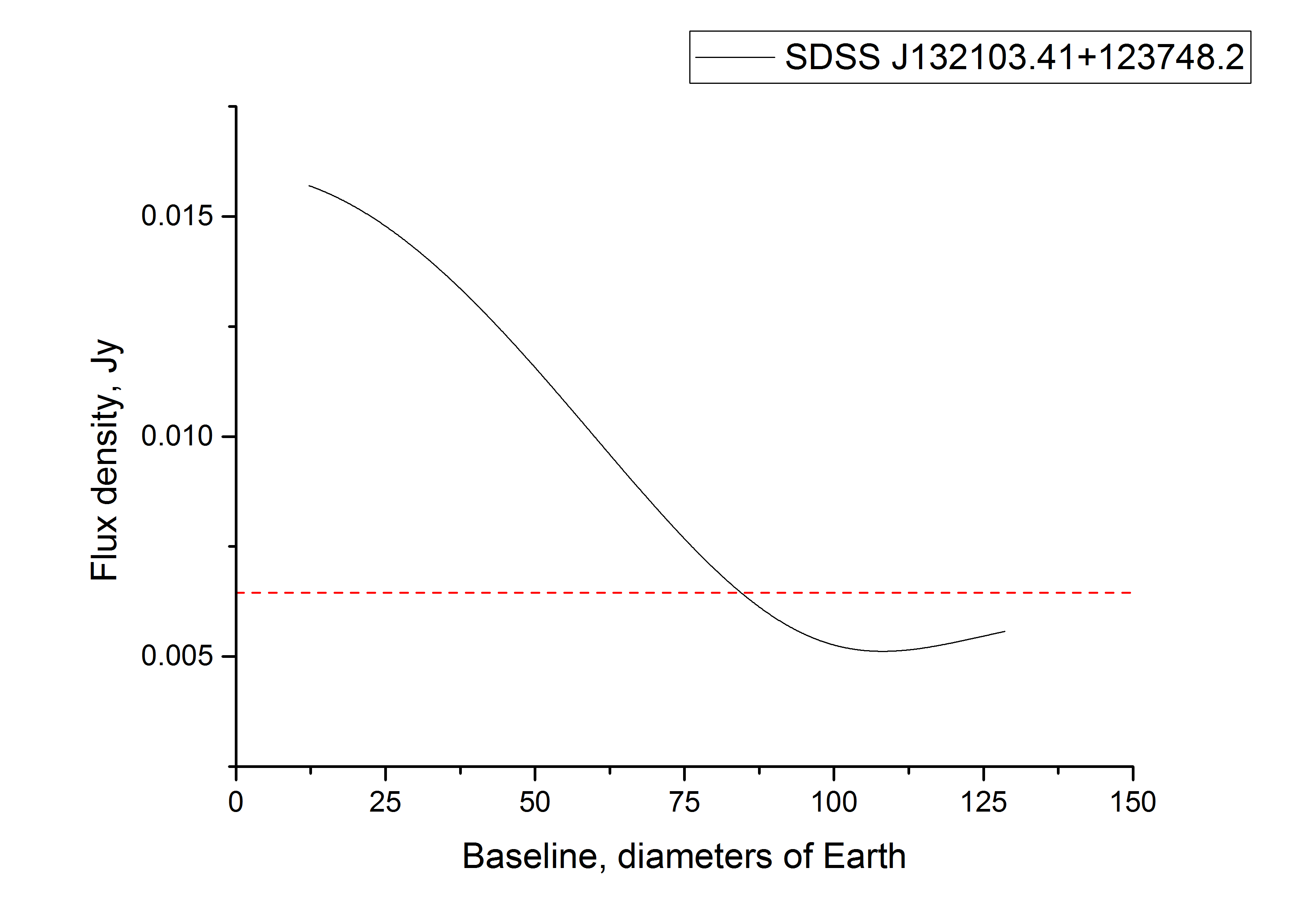}
\includegraphics[width=0.45\textwidth]{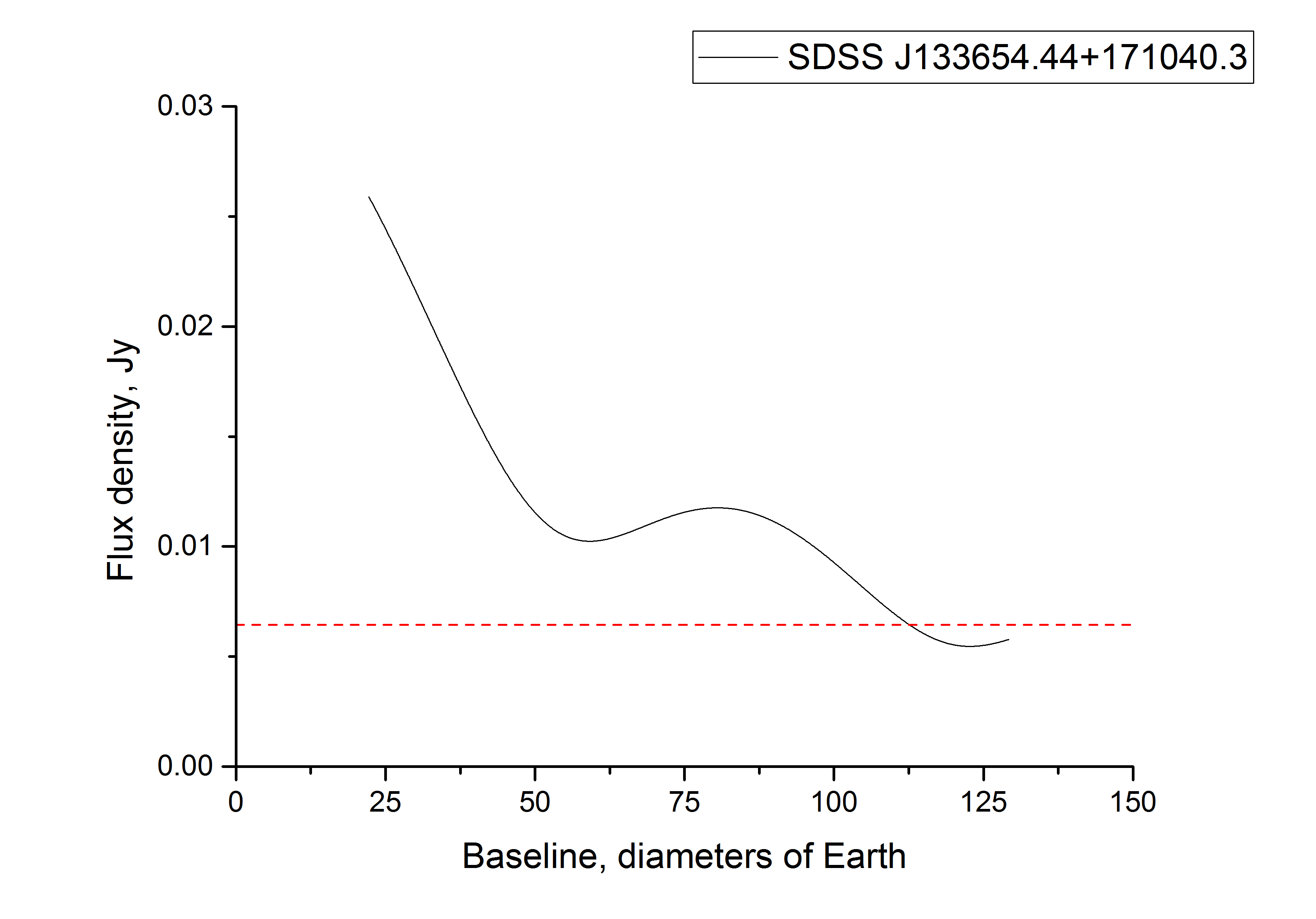}
\caption{Averaged visibility function for acceptable values of base projection for the sources from Table~1. Part~2.}
\end{center}
\label{flux2}
\end{figure}
\begin{figure}
\begin{center}
\includegraphics[width=0.45\textwidth]{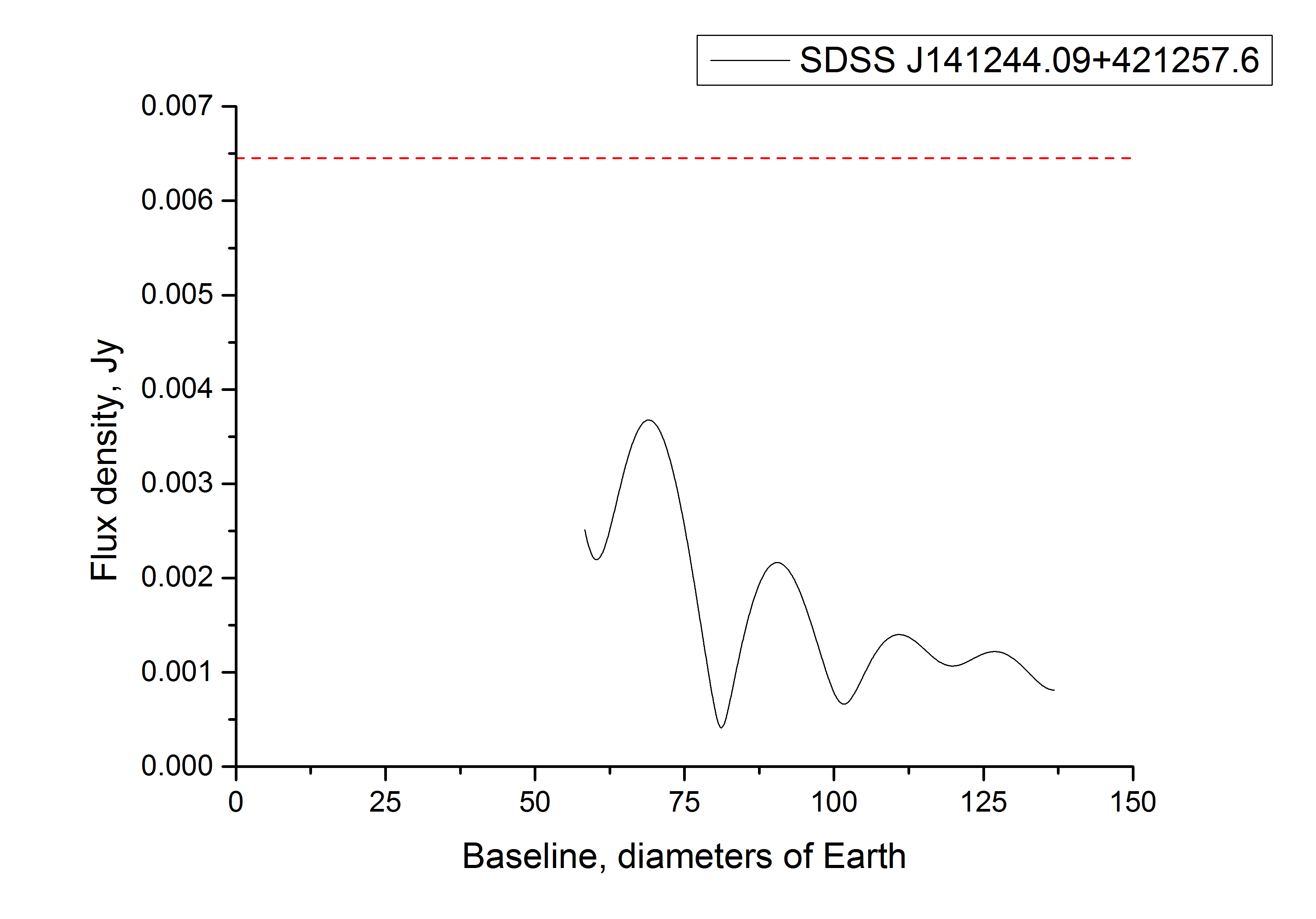}
\includegraphics[width=0.45\textwidth]{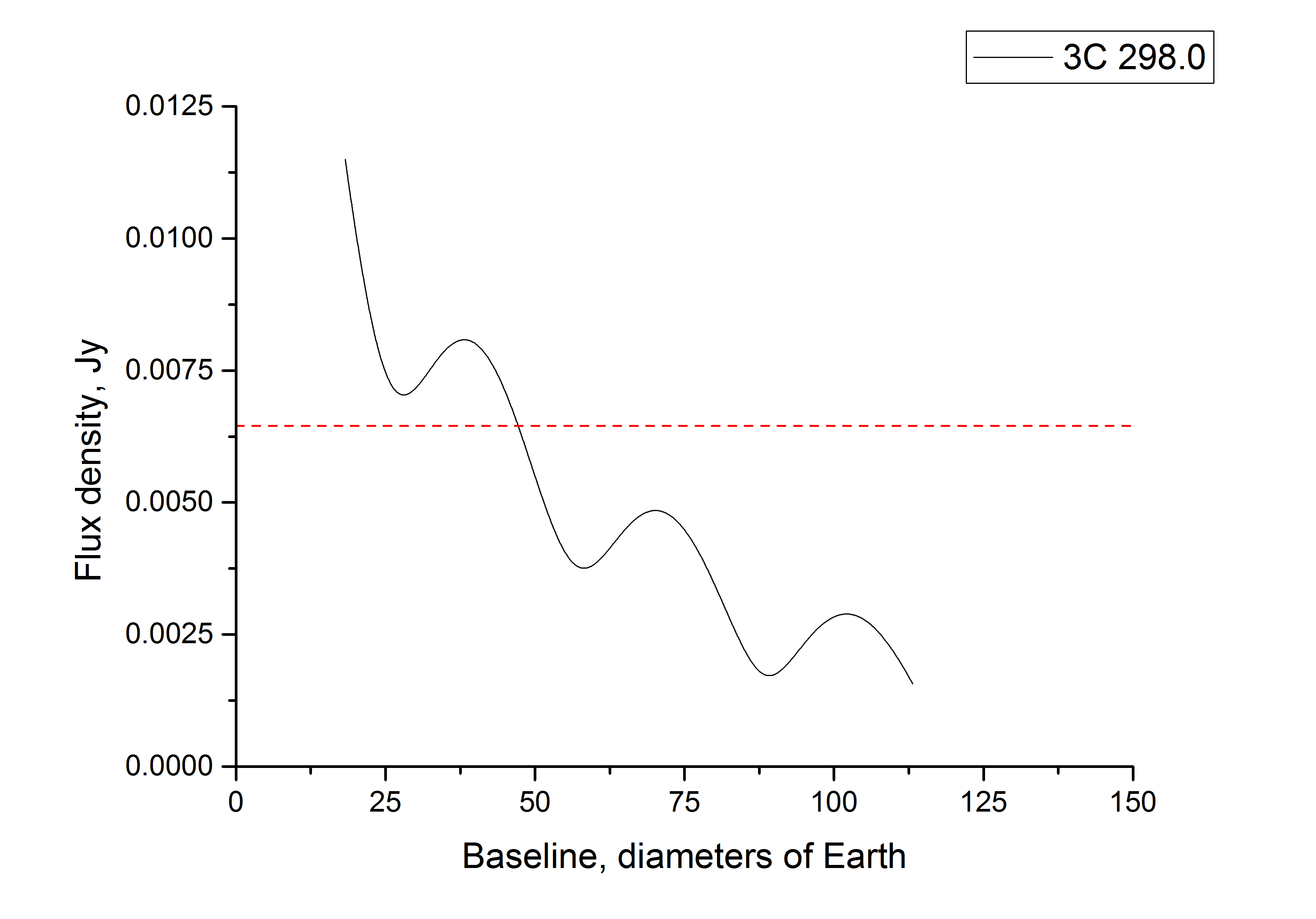}\\
\includegraphics[width=0.45\textwidth]{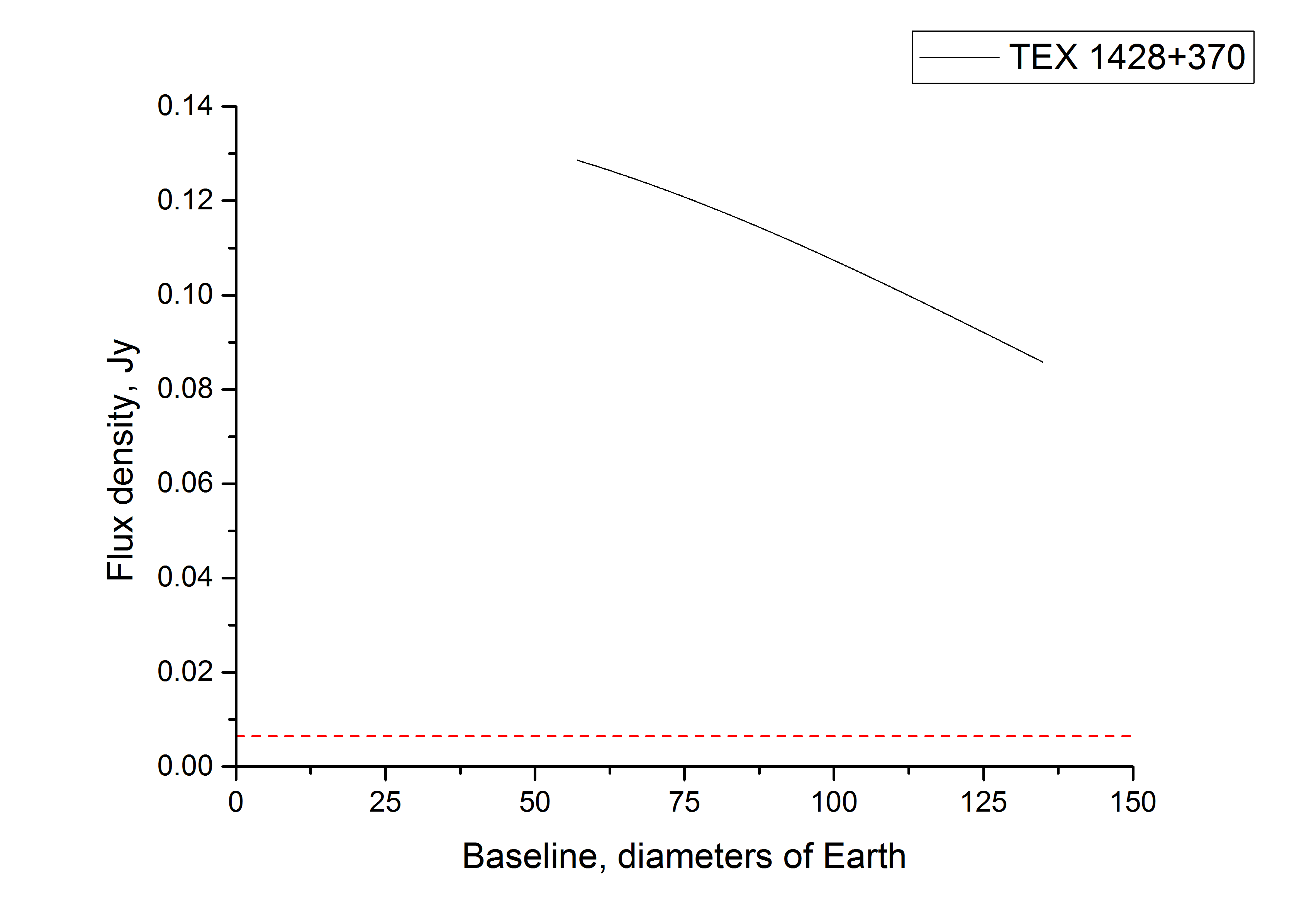}
\includegraphics[width=0.45\textwidth]{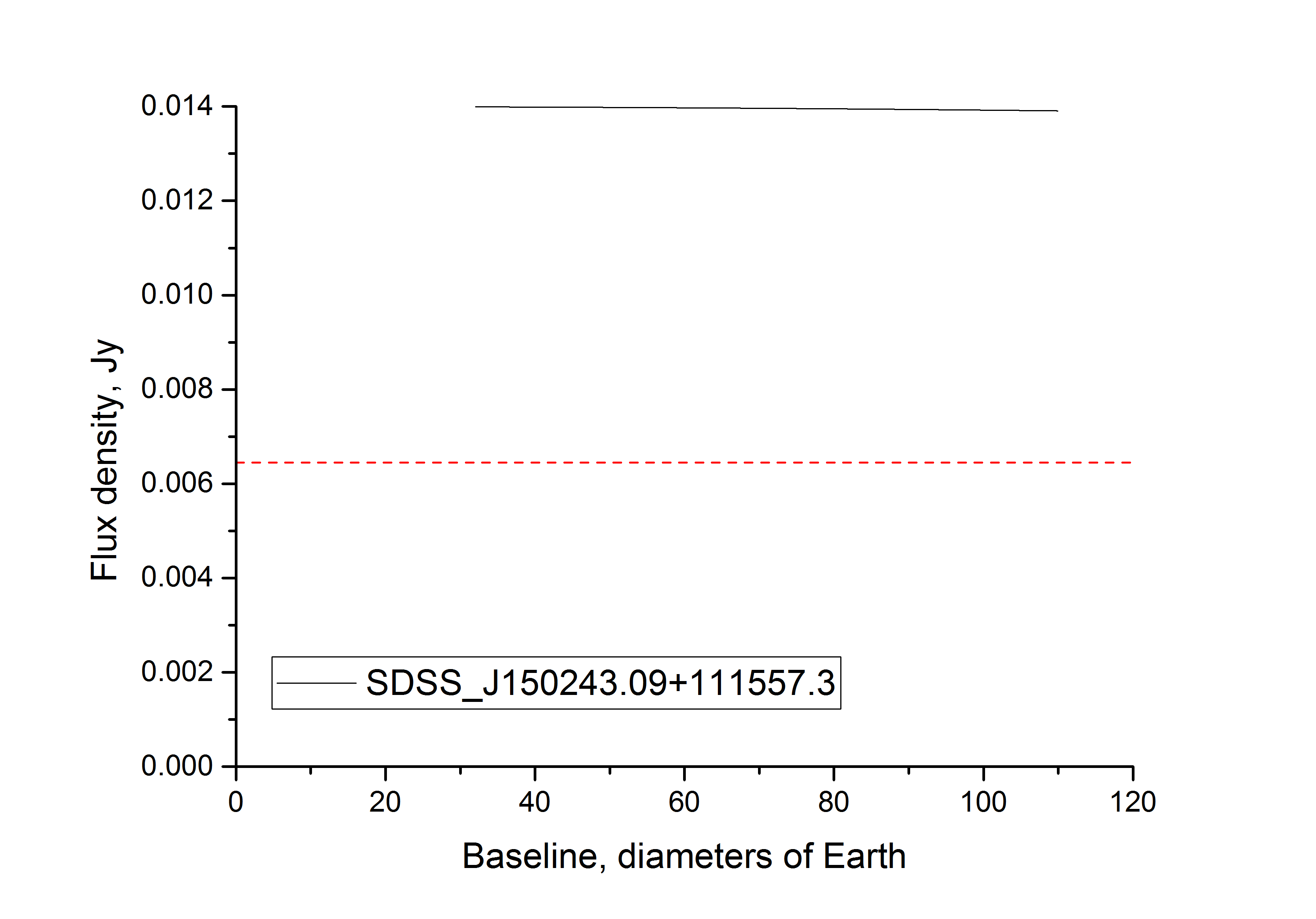}\\
\includegraphics[width=0.45\textwidth]{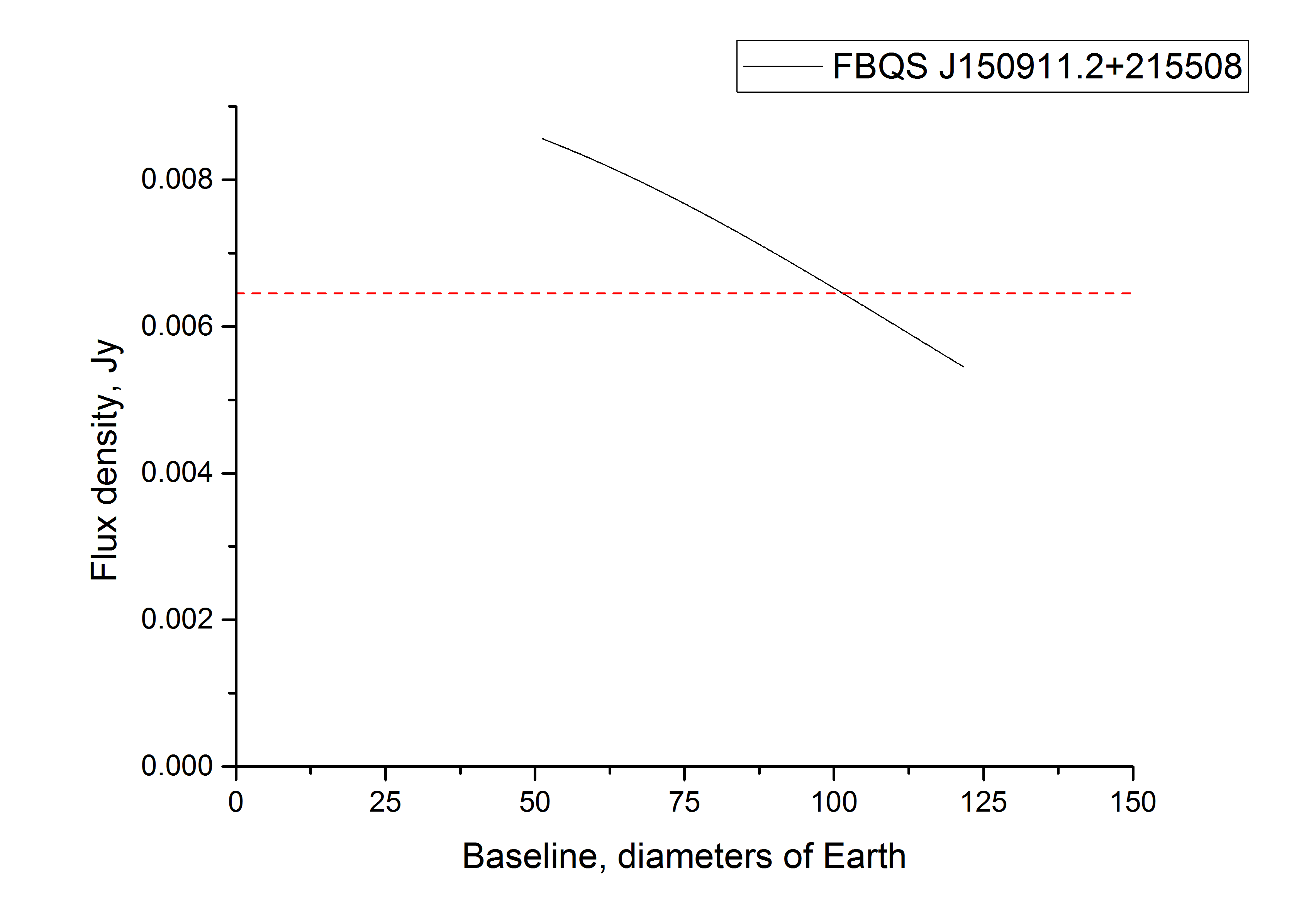}
\includegraphics[width=0.45\textwidth]{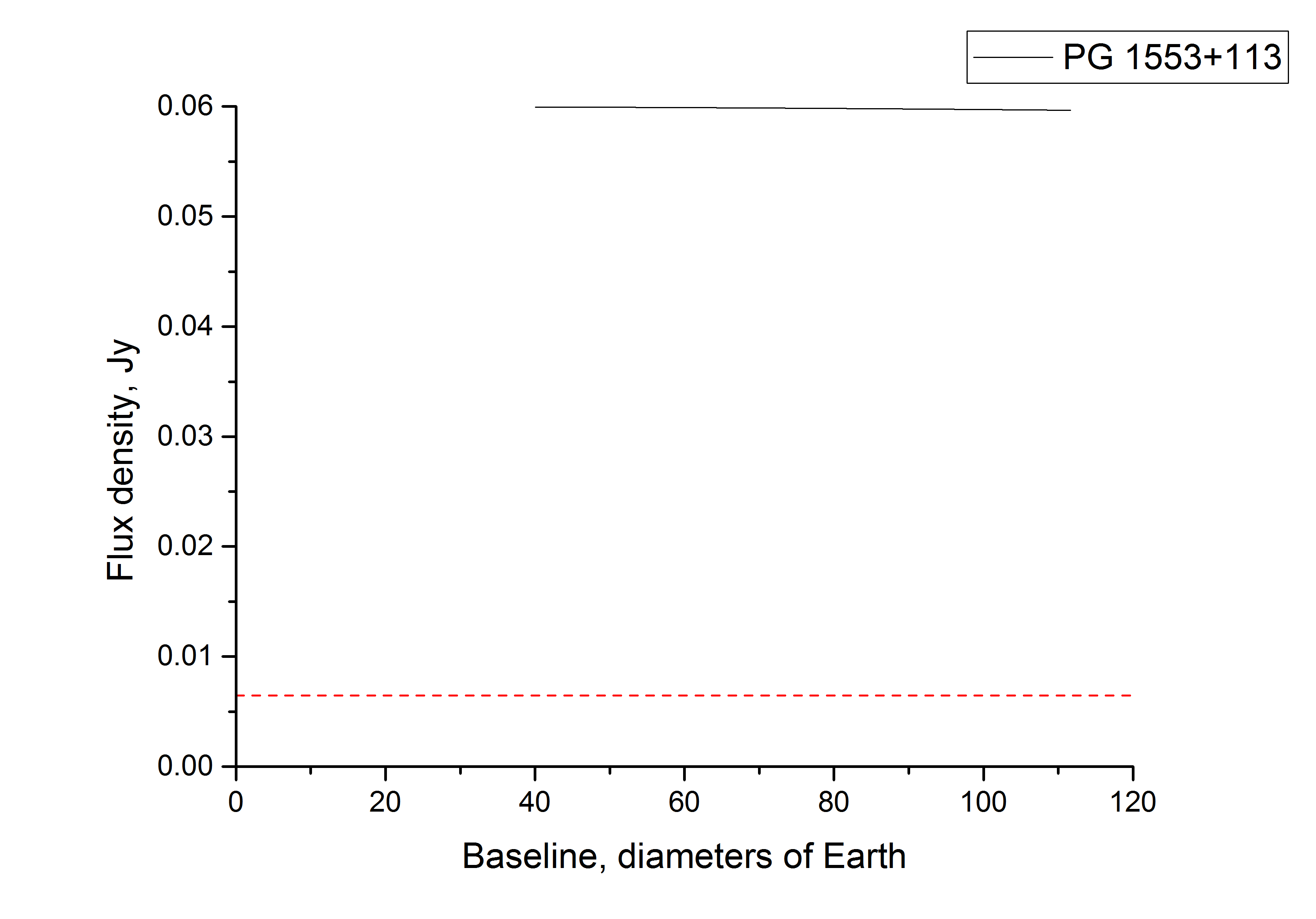}\\
\includegraphics[width=0.45\textwidth]{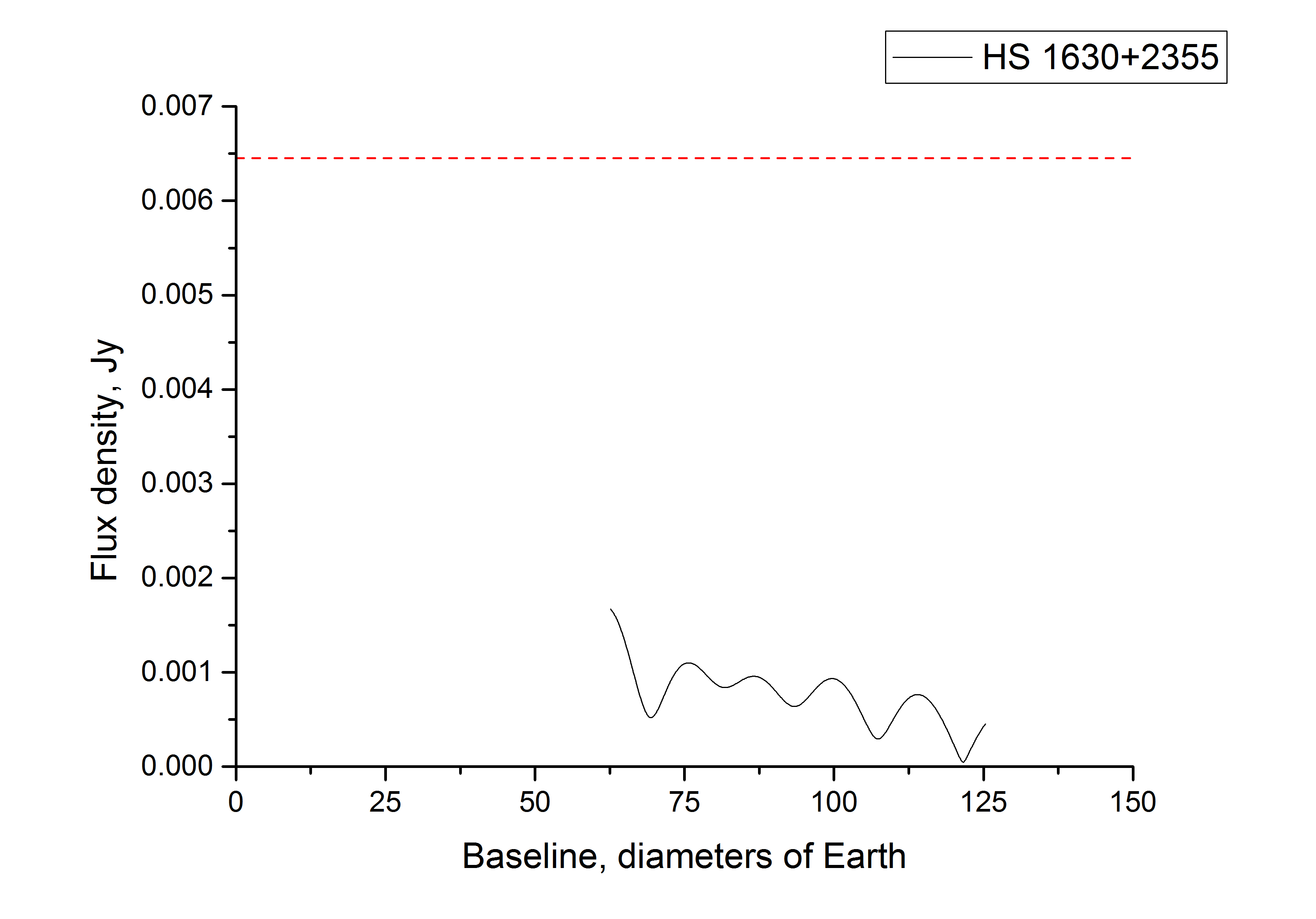}
\includegraphics[width=0.45\textwidth]{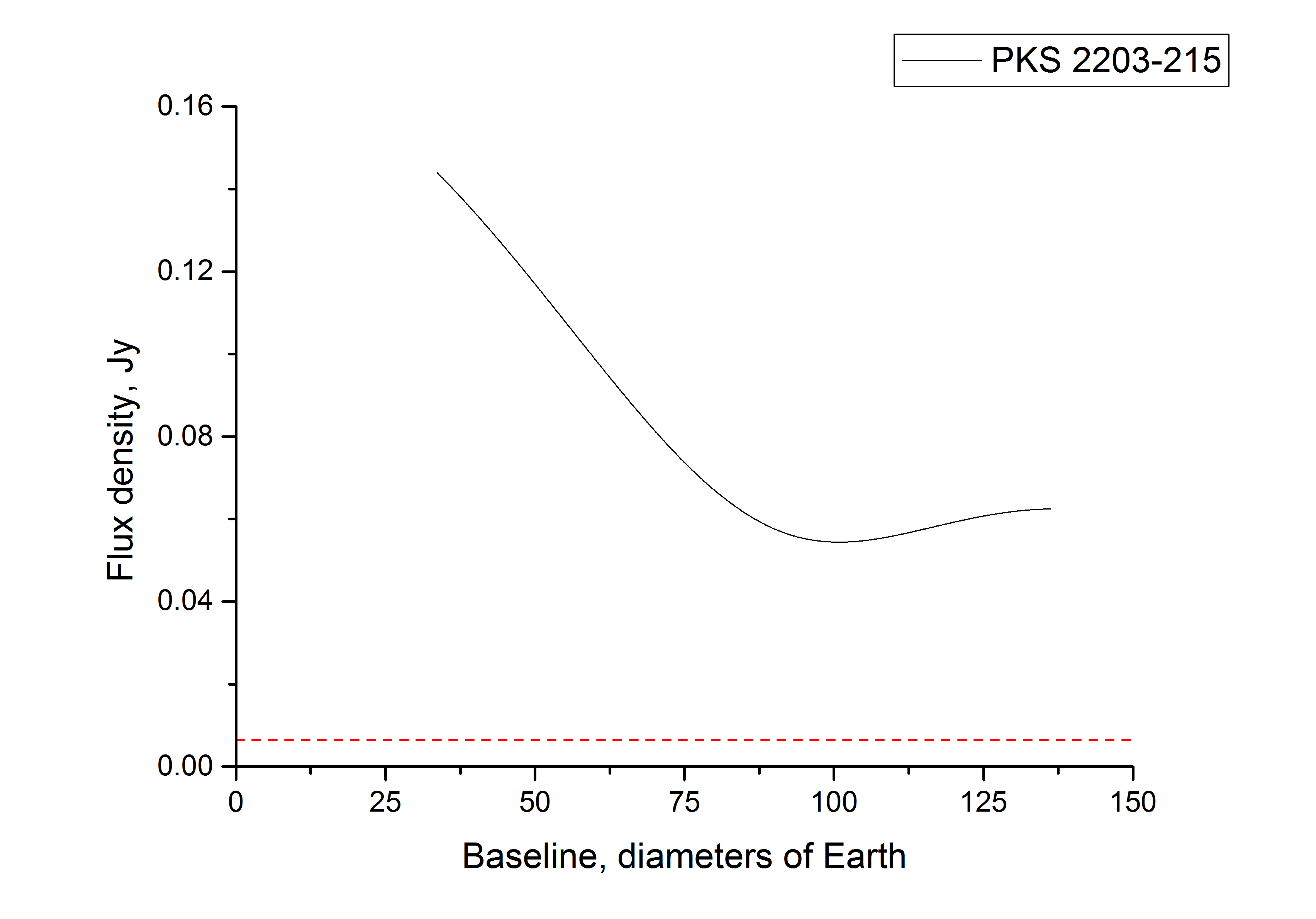} 
\caption{Averaged visibility function for acceptable values of base projection for the sources from Table~1. Part~3.}
\end{center}
\label{flux3}
\end{figure}

Table~2 summarizes the results of the analysis. It includes the sources from the  Table~1 for which there is an interval of base projections at which the amplitude of the visibility function is above the threshold value. Table~2 presents the name of the source, its right ascension, $\alpha$, and declination, $\delta$, at epoch J2000, the modeled flux $F_{ANN}$, expressed in Jy, and the angular size of the shadow of the more massive SMBH, $\theta$ (defined also as in \cite{catalmikh}), measured in arc microseconds, $\mu s$, and the angular distance between the components of the binary system, $d$, expressed $\mu s$ too. 

\begin{table}[t] 
\caption{SMBBH candidate meeting the selection criterion 
}
\begin{center} 
\begin{tabular}{|l|l|l|c|c|c|} 
\hline
Source name & $\alpha$, & $\delta$, & $F_{ANN}$,             & $\theta$, & $d$, \\
         & h\;m\;s   & ${}^o\; \prime\;\prime\prime$ & Jy & $\mu$s & $\mu$s \\
\hline
CSO 0402+379             & 04 05 09.3 & +38 03 32.2 & 0.043 & 0.068  & 6580  \\
FBQS J081740.1+232731    & 08 17 40.2 & +23 27 32.0 & 0.027 & 0.220  & 1.375 \\ 
BZQ J0842+4525           & 08 42 15.3 & +45 25 45.0 & 0.050 & 0.173  & 1.385 \\
OJ 287                   & 08 54 48.9 & +20 06 30.6 & 2.72  & 1.935  & 11.96 \\
MCG +11-11-032           & 08 55 12.5 & +64 23 45.6 & 0.23  & 0.067  & 4.82  \\
SBS 0924+606B            & 09 28 37.98& +60 25 21.0 & 0.019 & 0.087  & 9.69  \\
SDSS J102349.38+522151.2 & 10 23 49.5 & +52 21 51.8 & 0.027 & 0.237  & 1.717 \\
SDSS J124044.49+231045.8 & 12 40 44.5 & +23 10 46.1 & 0.014 & 0.058  & 1.072 \\
BZQ J1305-1033           & 13 05 33.0 & -10 33 19.1 & 0.37  & 0.035  & 1.794 \\
SDSS J132103.41+123748.2 & 13 21 03.4 & +12 37 48.1 & 0.016 & 0.055  & 1.0935 \\
SDSS J133654.44+171040.3 & 13 36 54.4 & +17 10 40.8 & 0.032 & 0.101  & 0.935 \\
3C 298.0                 & 14 19 08.2 & +06 28 35.1 & 0.022 & 0.213  & 1.498 \\
TEX 1428+370             & 14 30 40.6 & +36 49 03.9 & 0.14  & 0.025  & 0.319 \\
SDSS J150243.09+111557.3 & 15 02 43.1 & +11 15 57.3 & 0.014 & 0.01   & 25500 \\
FBQS J150911.2+215508    & 15 09 11.2 & +21 55 08.8 & 0.0094 & 0.029 & 0.411 \\
PG 1553+113              & 15 55 43.0 & +11 11 24.4 & 0.06  & 0.01   & 0.73  \\
PKS 2203-215             & 22 06 41.4 & -21 19 40.5 & 0.17  & 0.059  & 0.602 \\
 \hline
 \end{tabular}
 \end{center} 
\end{table} 

\section{Results} 

We have analysed the available list of candidates for binary SMBHs (SMBBHs). The list was compiled on the basis of available data on the variability in the optical range or the type of emission spectrum. In order to estimate the radiation flux at 240 GHz, we constructed an artificial neural network. For those SMBBH candidates for which this procedure appeared possible, the criterion of the possibility of observing the source at the Millimetron Space Observatory (MSO) was checked. 

The result of the study is presented in Table~2. It represents a list of 17 candidates the duality of which can be confirmed (or disproved) by observation with space-ground interferometer with an orbit and sensitivity at 240 GHz, as the planned MSO. 

The table with SMBBH candidates may be extended after additional observations or studies, as the main reason why the list is much shorter than the initial number of SMBBH candidates is the lack of observational data at a frequency of 240 GHz (at which the modeling was carried out), mainly at lower frequencies. 

\section{Discussion}

Since SMBBHs are an important but insufficiently studied stage in the evolution of SMBHs, their search and observations are very important. At the same time, most of the existing methods for their search (see \cite{review}), aimed at identifying candidates for SMBBHs, require confirmation, i.e., additional studies and direct space-VLBI observations.

As mentioned in the Introduction, the probability that there is a pair of gravitationally linked pairs of black holes is $\sim 10^{-3}$. In analyzing the survey CATALINA was able to select just over a hundred candidates for the binary SMBH out of a quarter of a million active globular nuclei, which is $\sim 4\times 10^{-4}$ and does not contradict the theoretical estimate.

The growing interest in SMBBH is also related to the planned space gravitational-wave observatories \cite{caprini2018}, the development of which is currently being widely discussed. However, results in this area can only be expected in the distant future. Thus, it is in radio interferometric observations that the first  SMBBHs may be discovered. This is due to the fact that SMBBH candidates are selected on the basis of properties that may be due not only to the duality of the SMBH but also to other physical reasons. In this context, it is particularly important to construct as complete as possible a catalog for interferometric observations in which the binary nature can be unambiguously established.

\section{Acknowledgement}

The authors are grateful to P.B. Ivanov for constructive suggestions and to A.G. Rudnitsky and M.A. Shchurov for help in performing calculations.

\end{document}